\documentclass[preprint,twocolumn,5p]{elsarticle}

\usepackage{lineno,hyperref,gensymb,array,eurosym,multirow,longtable}
{}
{}
{}
\usepackage{subfig}
\usepackage{amsmath}
\usepackage{amsfonts}

\usepackage{transparent}
\usepackage{booktabs}
\usepackage{multicol}
\usepackage{multirow}
\usepackage{hyperref}
\usepackage{float}
\usepackage{booktabs}
\usepackage[utf8]{inputenc}
\usepackage{color}
\definecolor{ultramarine}{RGB}{0,32,96}
\usepackage{lineno}
\biboptions{numbers,sort&compress}

\journal{Int. J. of Electrical Power \& Energy Systems \; {\url{https://doi.org/10.1016/j.ijepes.2020.106044}}}

\begin{document}
\nolinenumbers
\begin{frontmatter}

  \title{Frequency Control Analysis based on Unit Commitment Schemes with High Wind Power Integration: a Spanish Isolated Power System Case Study}

\author[mymainaddress]{Ana Fern\'{a}ndez-Guillam\'{o}n\corref{mycorrespondingauthor}}
\ead{ana.fernandez@upct.es}

\author[mysecondaryaddress]{Jos\'{e} Ignacio Saras\'{u}a}
\ead{joseignacio.sarasua@upm.es}

\author[mysecondaryaddress]{Manuel Chazarra}
\ead{manuel.chazarra@upm.es}

\author[mytertiaryaddress]{Antonio Vigueras-Rodr\'{i}guez}
\ead{avigueras.rodriguez@upct.es}


\author[mysecondaryaddress]{Daniel Fern\'{a}ndez-Mu\~noz}
\ead{daniel.fernandezm@upm.es}

\author[mymainaddress]{\'{A}ngel Molina-Garc\'{i}a\corref{cor1}}
\ead{angel.molina@upct.es}
\cortext[cor1]{Corresponding author}

\address[mymainaddress]{Dept. of Electrical Engineering, Universidad Polit\'{e}cnica de Cartagena, 30202 Cartagena, Spain}
\address[mysecondaryaddress]{Dept. of Hydraulic, Energy and Environmental Engineering, Universidad Polit\'{e}cnica de Madrid, 28040, Spain}
\address[mytertiaryaddress]{Dept. of Civil Engineering, Universidad Polit\'{e}cnica de Cartagena, 30202 Cartagena, Spain}

\begin{abstract}
  The relevant integration of wind power into the grid has involved a
  remarkable impact on power system operation, mainly in terms of
  security and reliability due to the inherent loss of the rotational
  inertia as a consequence of such new generation units decoupled from
  the grid. {\textcolor{black}{In these weak scenarios, the contribution of wind turbines to frequency control is considered as a suitable solution to improve system stability.  
  With regard to frequency response analysis and grid
  stability, most contributions introduce wind control discuss generation tripping for
  isolated power systems under arbitrary power imbalance
  conditions. Frequency response is then analyzed for hypothetical
  imbalances usually ranged between 5\% and 20\%, and assuming
  averaged energy schedule scenarios. In this paper, a more realistic
  framework is proposed to evaluate frequency
  deviations by including high wind power integration. With this aim,
  unit commitment schemes and frequency load shedding are considered
  in this work for frequency response analysis under high wind power
  penetration. The Gran Canaria Island's isolated power system (Spain)
  is used for evaluation purposes. Results provide a variety of
  influences from wind frequency control depending not only on the
  wind power integration, but also the generation units under operation, the rotational inertia reductions as well as the available reserves from each resource, aspects that have not been addressed previously in the specific literature to evaluate frequency excursions under high wind power integration.}}
\end{abstract}


\end{frontmatter}

\section*{Nomenclature}
\begin{table}[h]
  \begin{flushleft}
    \resizebox{0.985\columnwidth}{!}{ 
      \begin{tabular}{ll} 
        $cc$ & Combined cycle (subscript)\\
        $ds$ & Diesel (subscript)\\
        $D$ & Load-frequency sensitivity parameter (damping factor)\\
       $f$ & Frequency \\
       $g$ & Gas (subscript)\\
       $K$ & Participation factor\\
       $H_{i}$ & Rotational inertia of synchronous generator $i$\\ 
       $h_{WT}$ & Number of wind turbines\\
       $P_{i}$ & Power supplied by generation units $i$\\
       $P_{d}$ & Power demand \\
       $P_{J}$ & Power of \emph{Jin\'{a}mar} power plant\\
       $P_{MPPT}$ & Maximum power point of wind turbines\\
       $P_{sp}$ & Set-point active power of wind turbines\\
       $P_{T}$ & Power of \emph{Barranco de Tirijana} power plant\\
       $P_{w}$ & Power of wind power plants \\
       $RR$ & Regulation effort\\
      \end{tabular}
    }
  \end{flushleft}
\end{table}

\begin{table}[h]
  \begin{flushleft}
    \resizebox{0.685\columnwidth}{!}{ 
      \begin{tabular}{ll} 
       $s$ & Steam (subscript)\\
       $T_{m}(t)$ & Inertia of the power system\\
      $T_{u,th}$ & Secondary response time constant \\
      $v_{w}$ & Wind speed \\
       \hline
       AGC & Automatic generation control \\
       RES & Renewable energy sources \\
       MILP & Mixed integer linear programming\\
       RoCoF & Rate of change of frequency\\
       TSO & Transmission system operator\\
       UC & Unit commitment\\
       VSWT & Variable speed wint turbines\\
      \end{tabular}
    }
  \end{flushleft}
\end{table}

\section{Introduction} \label{sec.introduction} 

Conventional power plants with synchronous generators have
traditionally determined the inertia of power
systems~\cite{tielens16}. However, during the last decades, most
countries have promoted large-scale integration of Renewable Energy
Sources (RES)~\cite{zhang17,fernandez19power}. RES are usually not
connected to the grid through synchronous machines, but through power
electronic converters 
electrically decoupled from the grid~\cite{junyent15,tian16}. As a
consequence, by increasing the amount of RES and replacing synchronous
conventional units, the effective rotational inertia of power systems
can be significantly
reduced~\cite{akhtar15,yang18,fernandez19analysis}. Actually, Albadi
\emph{et al.} consider that the impact of RES on power systems mainly
depends on the RES integration and the system inertia~\cite{albadi10},
being the RES negative effects more severe in isolated power
systems~\cite{martinez18}. Among the different RES, wind power is the
most developed and relatively mature technology
{\cite{BREEZE2014223}}, especially variable speed wind
turbines~(VSWT)~\cite{edrah15,syahputra16,artigao18,cardozo18,li18}. Indeed,
Toulabi \emph{et al.} affirm that the participation of wind power into
frequency control services becomes inevitable due to the relevant
integration of such resource~\cite{toulabi17}.

Power imbalances between generation and demand can occur, among
others, due to the loss of power
generators~\cite{sokoler16}. Actually, this loss of power generators
can be the most severe contingency in case it is the largest
one~\cite{zhang18}. These imbalances cause frequency fluctuations, and
subsequently the grid becomes unstable, even leading to
black-outs~\cite{khalghani16,marzband16}. Hence, the frequency control
services are playing an essential role for secure and reliable power
systems~\cite{ozer15}. Moreover, frequency stability is the most
critical issue in isolated power systems due to their low rotational
inertia~\cite{aghamohammadi14,jiang15,jiang16}. Frequency control has
a hierarchical structure, and in Europe is usually organized up to
five layers: $(i)$~frequency containment, $(ii)$~imbalance netting,
$(iii,iv)$~frequency restoration (automatic and/or manual) and
$(v)$~replacement, from fast to slow timescales~\cite{entsoe_europe}.

According to the specific literature, several studies have proposed
wind frequency control approaches. However, authors notice that in
those contributions: $(i)$ energy schedule scenarios considered are
usually arbitrary and unrealistic, without considering a unit
commitment~(UC) scheme and individual generation
units~\cite{keung09,el11,mahto17,abazari18,alsharafi18}; $(ii)$ the
power imbalance is usually taken as a fixed random value (between 3
and 20\%), excluding the $N-1$
criterion~\cite{ma10,wang13,kang16,ochoa18}; $(iii)$ load shedding is
not taken into account in the frequency response analysis
{~\cite{bevrani11,ma18,8667397}}; and $(iv)$~only a few wind power
integration scenarios are commonly analyzed to evaluate the wind
frequency controller ---usually one or two different
scenarios---~\cite{zhang12,mi16,bao18,chen18}. As a
consequence, simulations can address unrealistic and inaccurate results. For
instance, recent studies considering two wind integration share rates
provide different ---and even opposite--- conclusions regarding
frequency nadir and RoCoF (Rate of Change of Frequency): some authors
conclude that these parameters can improve~\cite{wang13,wilches15},
others that they could get worse~\cite{alsharafi18,aziz18} or even be
similar~\cite{kang16} as wind penetration increases.

By considering previous contributions, the aim of this paper is to
analyze the frequency response of an isolated power system with high
integration of wind power generation including wind frequency control
and load shedding. These energy schedule scenarios are determined by a
UC model, taking into account some technical and economical
constraints~\cite{farrokhabadi18} and guaranteeing the frequency
system recovery after the largest power plant outage ($N-1$
criterion)~\cite{teng16}. A realistic load shedding program is also
included, as well as wind frequency control. With this aim, our study
is carried out in Gran Canaria Island power system, in the Canary
island archipelago~(Spain), where the wind power integration has
increased from 90 to 180~MW in the last two years. Moreover, in the
Canary island archipelago, more than 200 loss of generation events per
year were registered between 2005 and 2010. In fact, the number of
this kind of incidents even surpassed 300 per year, subsequently
suffering from the activation of the load shedding programs
\cite{padron2015reducing}. This analysis can be extended to other
isolated power systems with relevant wind energy potential, such as
Madagascar~\cite{praene17} or Japan~\cite{meti15}. The main
contributions of this paper can be thus summarized as follows:
\begin{itemize}
\item Evaluation of wind frequency control responses,
  including load shedding and rotational inertia changes from
  realistic operation conditions under generation unit
  tripping.
\item Analysis of frequency deviations (nadir, RoCoF) in
  isolated power systems with high penetration of wind power,
  using energy schedules and unit comments obtained from an
  optimization model.
\end{itemize}

The rest of the paper is organized as follows:
Section~\ref{sec.power_system_description} describes the Gran Canaria
power system and the generation scheduling process. The power system
model, including both optimization and dynamic models, are described
in Section~\ref{sec.power_system_model}. Simulation results are
analyzed and discussed in Section~\ref{sec.results}. Finally,
Section~\ref{sec.conclusions} outlines the main conclusions of the
paper.

\section{Power System and Generation Scheduling Process}\label{sec.power_system_description}

\subsection{Preliminaries}{\label{sec.power_system_general_overview}}

Different frequency analysis studies have been carried out by authors
based on specific power systems. For instance, Zerket \emph{et
  al.}. considered a modified Nordic 32-bus test system
\cite{zertek12}; in \cite{nazari2014distributed}, the power system of
Flores Island, and the electric power system of Sao Miguel Island were
used; Moghadam \emph{et al.}. focused on the power system of Ireland
\cite{moghadam2014distributed}; in \cite{wang2017system}, the
Singapore power system was used; Pradhan \emph{et al.} tested the
three-area New England system \cite{pradhan2019online}; and
\cite{sarasua2019analysis} considered the Spanish isolated power
system located in El Hierro Island. In this paper, authors have
focused on Gran Canaria Island (Spain), where the wind power
integration has increased from 90 to 180 MW in the last two years.

Gran Canaria Island belongs to Canarian archipelago,
  one of the outermost regions of the European Union. Canarian
  archipelago is located in the north-west of the African
  Continent. From the energy point of view, Gran Canaria Island is an
  isolated power system. Traditionally, Gran Canaria Island's generation has
  been exclusively associated with fossil fuels: diesel, steam, gas
  and combined cycle units from two different power plants:
  \emph{Jin\'{a}mar power plant} and \emph{Barranco de Tirijana power
    plant}. However, this fossil fuel dependence has involved an
  important economic and environmental drawback. To overcome theses
  problems, the Canary Government promoted the installation of wind
  power plants in the 90's, accounting for 70~MW in 2002. In the
  following decade, the installation of wind power plants stopped
  around 95~MW and, since 2015, wind power capacity has been doubled,
  nearly reaching 180~MW.

\begin{figure*}[tbp]
  \centering 
  \includegraphics[width=0.75\linewidth]{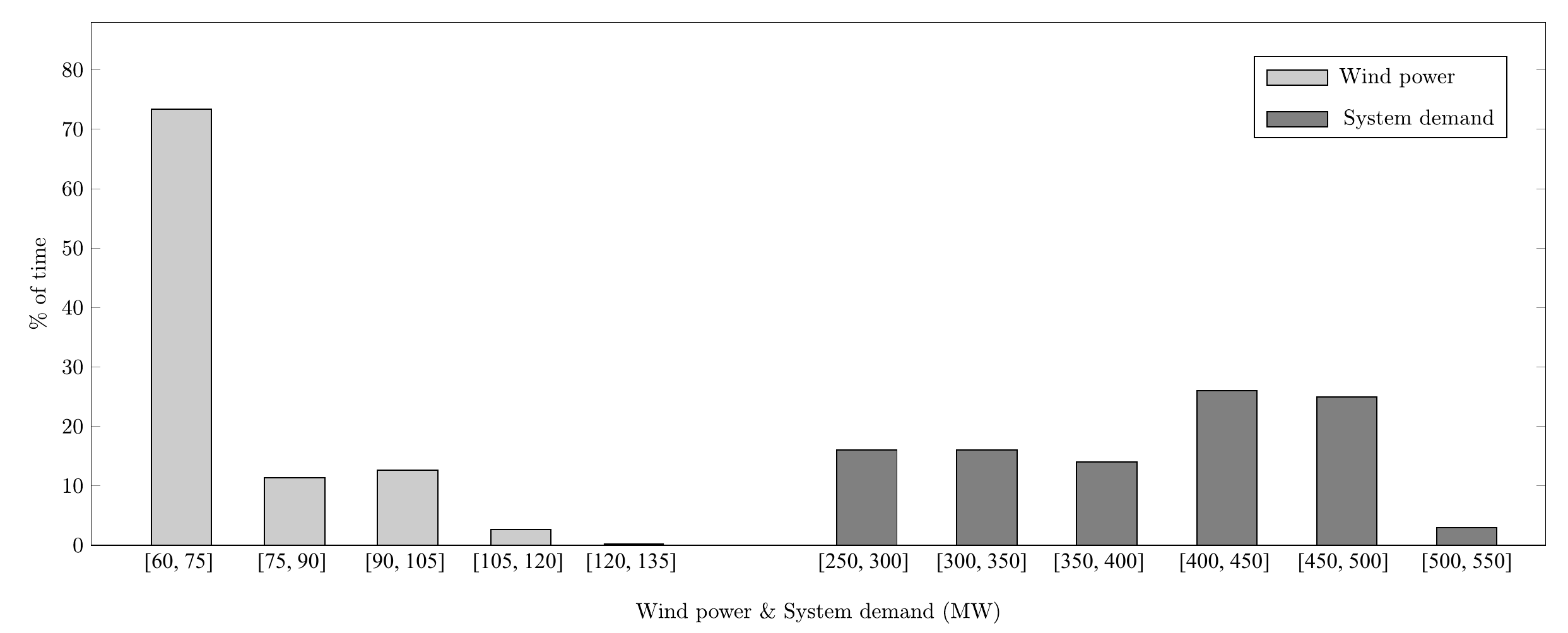}
  \caption{{\textcolor{black}{Wind power generation and system demand distribution in Gran Canaria Island power system along 2018~\cite{ree}.}}}
  \label{fig.demand}
\end{figure*}

  Regarding the wind power generation and system demand in Gran
  Canaria Island along 2018, both are shown in
  Fig.~\ref{fig.demand}. The system demand is discretized for six
  different intervals, considering the lowest and highest demand of
  Gran Canaria Island. Wind power generation is discretized for five
  intervals. According to the ranges in the system demand and the wind
  power generation shown in Fig.~\ref{fig.demand}, thirty energy
  scenarios are proposed to analyse the frequency response of the
  system including wind frequency control. Each energy scenario is
  based on a pair demand-wind power generation as it is further
  described in Section~\ref{sec.results}.

\subsection{Generation Scheduling Process}
\begin{figure}[tbp]
  \centering
  \subfloat[Weekly and daily scheduling]{\includegraphics[width=\linewidth]{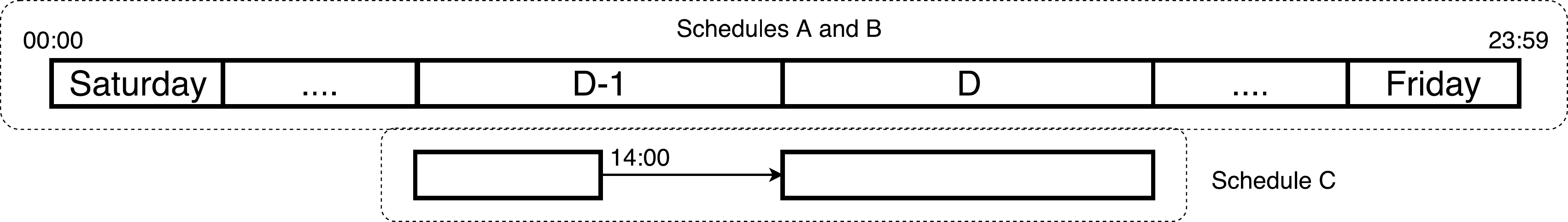}\label{fig.timeline}}\\
  \subfloat[Schedules A--C]{\includegraphics[width=0.4795\linewidth]{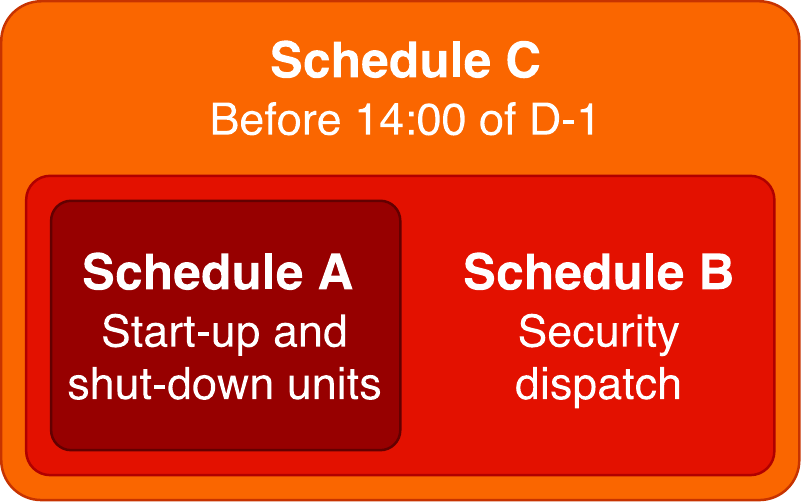}\label{fig.schedules}}
  \caption{\textcolor{black}{Generation scheduling of Gran Canaria Island power system}}
  \label{fig.schedule}
\end{figure}

{\label{sec.power_system_generation_scheduling}}
The generation scheduling of the Gran Canaria Island power system is
ruled in~\cite{miet12,miet15}. It is carried out by the Spanish
Transmission System Operator (TSO) according to the economic criterion
of variable costs of each power plant. The schedules are obtained
according to different time horizons: weekly or daily. Each energy
schedule depends on the previous time horizon and, subsequently,
weekly and daily schedules are required to determine the hourly
generation scheduling, which is used in the present paper.  An
overview of these schedules is summarized in
Fig.~\ref{fig.schedule}.

\begin{enumerate}
\item \emph{Weekly scheduling:} Estimation of the hourly start-up and
  shut-down decisions from each Saturday (00:00~h) to the following
  Friday (23:59~h). This initial generation schedule is determined
  following two steps: $(i)$~an economic dispatch is carried out to
  minimize the total variable costs to meet the net power system
  demand (i.e., the power system demand minus the renewable
  generation). The result of such economic dispatch includes both the
  hourly energy and the reserve schedules (labeled as
  {\texttt{Schedule-A}}). $(ii)$~an economic and security dispatch is
  determined taking into account the transmission lines and minimizing
  the total variable costs to support the net power system demand and
  a certain level of power quality. The result of this economic and
  security dispatch is also a hourly energy and reserve schedule
  (labeled as {\texttt{Schedule-B}}).
\item \emph{Daily scheduling:} Updates the {\texttt{Schedule-B}} of a
  certain day~$D$ from the updated available information of the power
  system: generation from suppliers, demand from consumers and the
  state of the transmission lines. The result of the daily scheduling
  in the day~$D$ is a new hourly energy and reserve schedule (labeled
  as {\texttt{Schedule-C}}). It is obtained before 14:00~h of the day~$D-1$. This
  {\texttt{Schedule-C}} is determined following a similar process as in the
  weekly scheduling: $(i)$~an economic dispatch is firstly carried out
  and $(ii)$~an economic and security dispatch is then
  calculated. The daily scheduling processes aims to
  minimize the total variable costs to meet the net power system
  demand with a minimum certain level of power quality.
\end{enumerate}

\section{Methodology: Unit Commitment and Frequency Model}\label{sec.power_system_model}

\begin{figure}[tbp]
	\centering
	\includegraphics[width=0.9\linewidth]{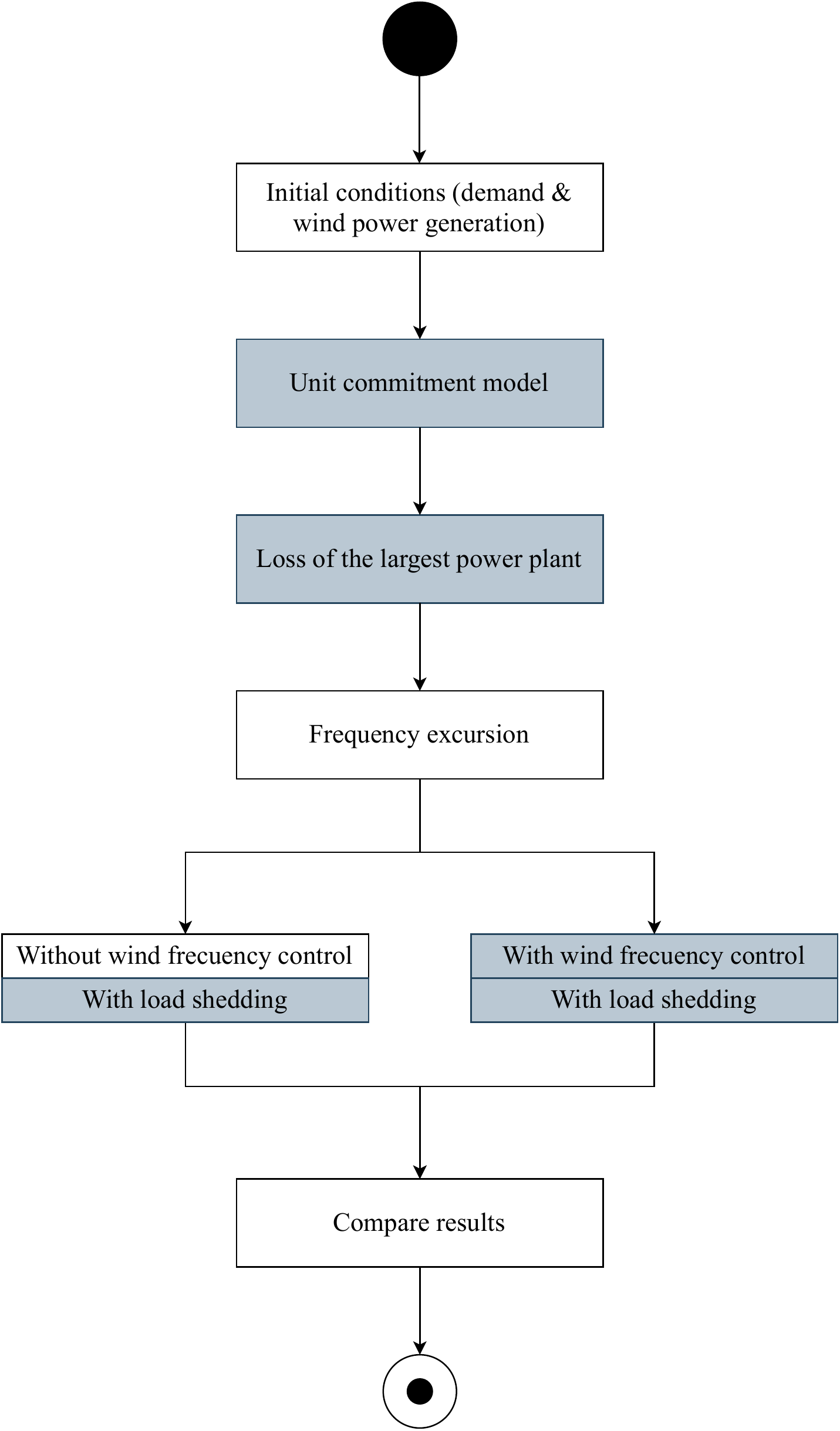}
	\caption{\textcolor{black}{Flow chart of the methodology used}}
	\label{fig.flow_chart}
\end{figure}

\begin{figure*}[tbp]
	\centering
	\includegraphics[width=0.595\linewidth]{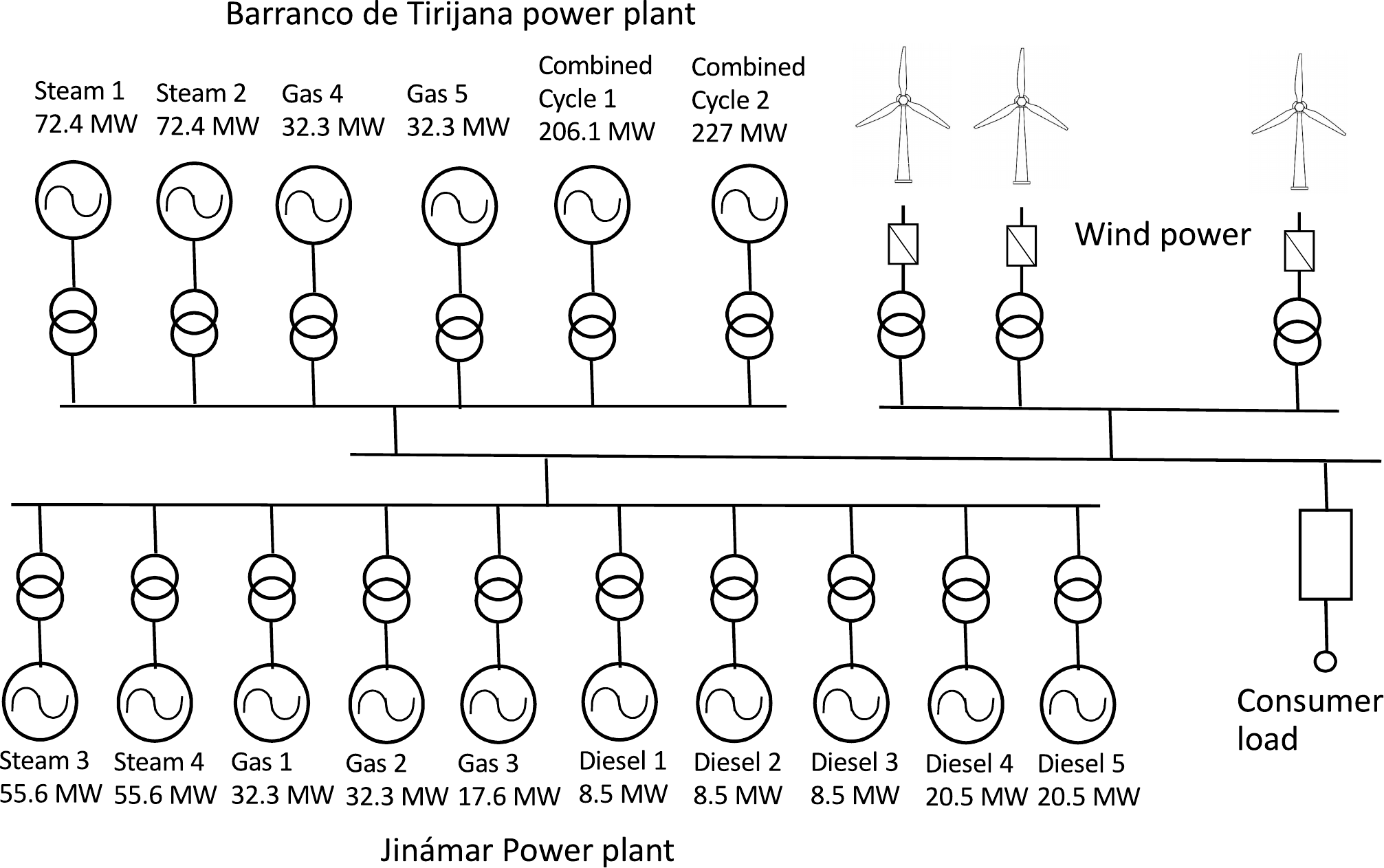}
	\caption{\textcolor{black}{Gran Canaria Island power system, simplified one-line diagram}}
	\label{fig.power_system}
\end{figure*}

Frequency deviations are analyzed according to possible generation
tripping and power system reserves by considering explicitly
individual generation units and technologies. With this aim, a series
of energy scenarios for each scenario of system demand and wind power
generation based on a real isolated power system (the Gran Canaria
Island) are estimated and evaluated accordingly, considering current
wind power integration percentages and load shedding
programs. Fig.~\ref{fig.flow_chart} shows the proposed methodology,
highlighting the novelties and differences presented in this paper
compared to other approaches focused on frequency analysis (i.e.,
carry out a UC to determine the energy scenarios, consider the loss of
the largest power plant as imbalance, and include load shedding with
and without wind frequency control). The following subsections
describe respectively the unit commitment model and frequency models
used in this work. Fig. \ref{fig.power_system} shows a simplified
scheme of the modelled Gran Canaria Island power system, where
conventional and wind power plants are depicted.

\subsection{Unit commitment model: creation of scenarios}\label{sec.optimization_model}

In order to analyze frequency deviations in the Gran Canaria power
system, a UC model is required to estimate the number of thermal units
connected to the grid for each generation mix scenario. These thermal
units remain unchanged during the subsequent frequency control
analysis. The UC model used in this paper has been recently proposed
by the authors in {\cite{FernandezMunoz2019}}, based on
{\cite{fernandez19}}, which is a deterministic thermal model based on
mixed integer linear programming (MILP). Other contributions focused
on probabilistic unit commitment can be also found in the specific
literature. In this way, {\cite{AZIZIPANAHABARGHOOEE2016634}} proposes
an optimal allocation of up/down spinning reserves under high
integration of wind power. The planning horizon of our model is
adapted to 24 hours with a time resolution of one hour consistent with
the approach used by the TSO for the next-day generation scheduling
{\cite{pezic13}}; and the hydropower technology is excluded from the
model formulation in order to be consistent with the generation mix of
the Gran Canaria power system. The model formulation is partially
based on {\cite{morales12}} which is, to the author's knowledge, the
most computationally efficient formulation available in the literature
when considering different types of start-up costs of thermal units.

The Gran Canaria power system is operated in a centralized manner by
the TSO in order to minimize the total system costs, according
to~\cite{miet15}. Therefore, the objective function of the UC model
here used consists in minimizing the start-up cost, the fuel cost, the
operation and maintenance cost and the wear and tear cost of all
thermal units. The optimal solution of the UC model is formed by the
hourly energy schedule of each thermal unit of the system, taking into
account their minimum on-line and off-line times. Among others, the
energy schedule is restricted by the following constraints. The
production-cost curve of each thermal unit is modeled as a piece-wise
linear function discretized by ten pieces. The number of pieces is
determined as a trade-off between the accuracy of the solution and the
computation time cost limits. In addition, the system demand and the
spinning reserve requirements must be fulfilled in each
hour. According to the P.O. SEIE 1 in~\cite{miet12}, the hourly
spinning reserve must be higher or equal to the maximum of the
following three values: $(i)$ the expected inter-hourly increase in
the system demand between two consecutive hours; $(ii)$ the most
likely wind power loss calculated by the TSO from historical data;
$(iii)$ the loss of the largest spinning generating unit in each
hour. It is important to bear in mind that there are two conceptual
differences between the formulation presented in~\cite{morales12} and
the one used here:
\begin{itemize}
\item The meaning of each start-up type of each thermal unit is
  different. In~\cite{morales12}, each start-up type corresponds to
  a different power trajectory of the thermal unit whereas the
  approach of the model here used is the following: each start-up
  type refers to the start-up cost as a function of the time that
  the unit has remained off-line since the previous shut-down. The
  start-up cost calculation and the involved parameters of the
  thermal units are defined in~\cite{miet15}.
\item For those thermal units that have a start-up process longer
  than one hour, a single output power trajectory ranging from zero
  to the unit's minimum output power is considered.
\end{itemize}
Further details of the UC model formulation can be found in \cite{FernandezMunoz2019}.

\subsection{\textcolor{black}{Frequency analysis model}}\label{sec.dynamic_model}

\subsubsection{General overview}\label{sec.dynamic_model_general_overview}

\begin{figure*}[tbp]
  \centering
  \includegraphics[width=.6\linewidth]{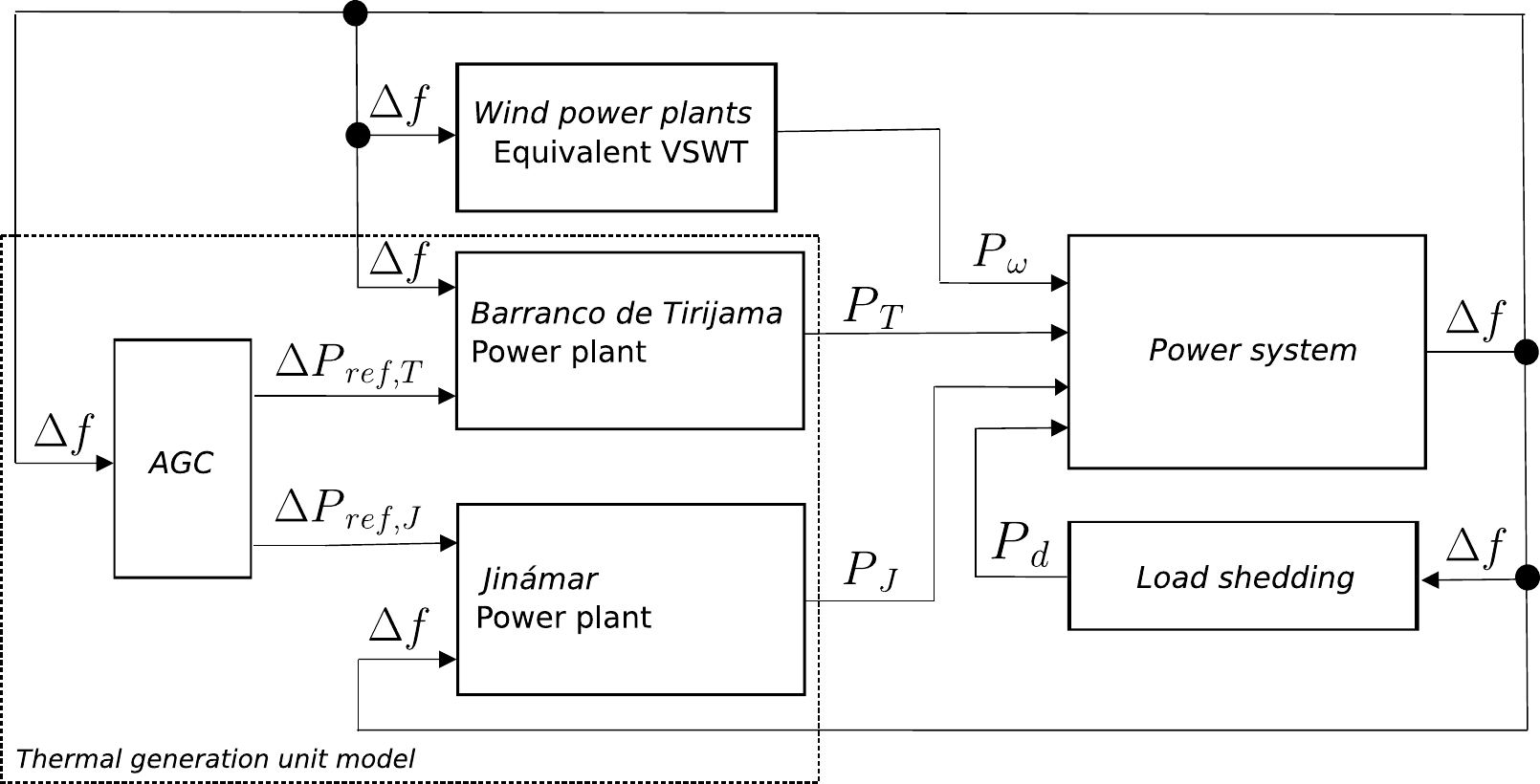}
  \caption{{\textcolor{black}{Frequency analysis model. Block diagram.}}}
  \label{fig.dynamic_model}
\end{figure*}

Frequency deviations in power systems are usually modeled by means of
an aggregated inertial model. This assumption has been successfully
applied to isolated power systems, as the Irish power system
\cite{mansoor00}. In this paper, frequency system variations are the
result of an imbalance between the supply-side (\emph{Barranco de
  Tirijana} power plant~$P_{T}$, \emph{Jin\'{a}mar} power
plant~$P_{J}$ and wind power plants~$P_{w}$, which are explicitly
considered for simulations) and the demand-side~$P_{d}$. A load
frequency sensitivity parameter $D$ is also included to model the load
sensitivity under frequency excursions~\cite{o14},
\begin{equation}\label{eq.swing}
f\,\dfrac{df}{dt}=\dfrac{1}{T_{m}(t)}\left(P_{T}+P_{J}+P_{w}-P_{d}-D\cdot\Delta f \right) ,
\end{equation}
being $T_{m}(t)$ the total inertia of the power system, which
corresponds to the equivalent addition of the rotational inertia of
all synchronous generators under operation conditions, in terms of the
system base,
\begin{equation}\label{eq.Tm}
T_{m} (t)= \sum_{m=1}^{4}2\;H_{s,m}+\sum_{q=1}^{5}2\;H_{g,q}+\sum_{k=1}^{5}2\;H_{ds,k}+\sum_{l=1}^{2}2\;H_{cc,l}\,.
\end{equation} 
Frequency and power variables also depend on time, but it is not
explicitly included to simplify the
expressions. Fig.~\ref{fig.dynamic_model} shows the general block
diagram of the proposed simulation model. It has been developed in
Matlab/Simulink. The block {\texttt{`Power system'}} in
Fig.~\ref{fig.dynamic_model} contains eq. ({\ref{eq.swing}}),
modeled with the corresponding block diagram. The power provided by
the power plants, $P_{T}$ and $P_{J}$ respectively, are the addition
of the power supplied by each thermal generation unit (steam~$s$,
gas~$g$, combined cycle~$cc$ and diesel~$ds$) under operating
conditions, expressed as follows:
\begin{equation}\label{eq.pt}
P_{T}=\sum_{m=1}^{2}P_{s,m}+\sum_{q=4}^{5}P_{g,q}+\sum_{l=1}^{2}P_{cc,l}\;,
\end{equation}
\begin{equation}\label{eq.pj}
P_{J}=\sum_{m=3}^{4}P_{s,m}+\sum_{q=1}^{3}P_{g,i}+\sum_{k=1}^{5}P_{ds,k}\;.
\end{equation}
The $k$, $l$, $m$ and $q$ indexes refer to the number of diesel,
combined cycle, steam and gas units of each power plant,
respectively. The proposed dynamic model depicted in
Fig.~\ref{fig.dynamic_model} is thus composed by the different
thermal units belonging to \emph{Barranco de Tirijana} and
\emph{Jin\'{a}mar} power plants ---explicitly considered in the
model---, as well as the wind power plants, the power system and the
power load of consumers. This load consumer block includes the load
shedding program, activated when the grid frequency exceed certain
thresholds. The dynamic response of each generation unit is simulated
according to the transfer functions discussed in Section
{\ref{sec.thermal_units}}.

\subsubsection{Thermal generation units}\label{sec.thermal_units} 

\begin{figure*}[tbp]
	\centering
	\includegraphics[width=0.755\linewidth]{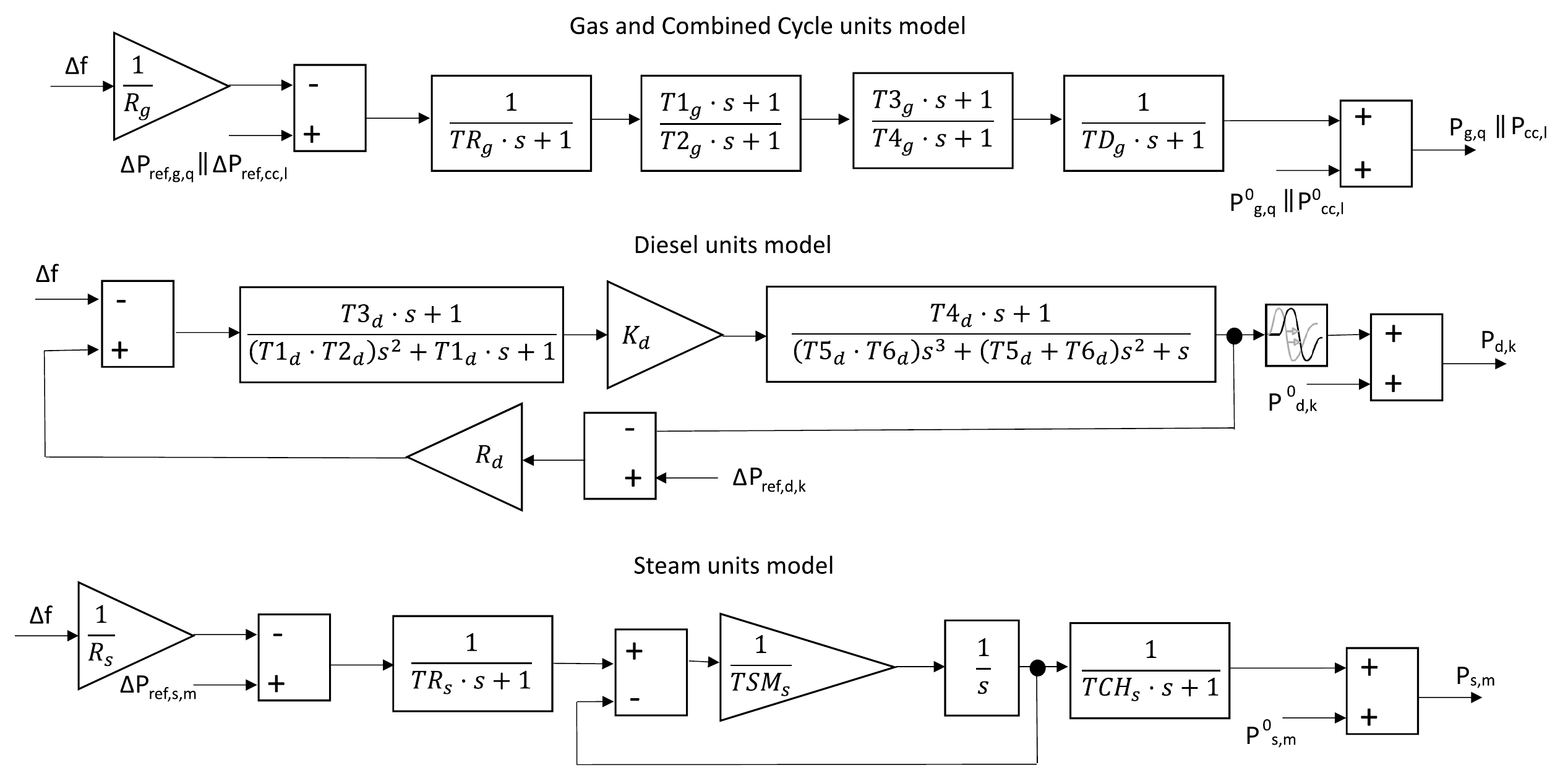}
	\caption{Block diagrams of thermal units models.}
	\label{fig.thermal_generation_models}
\end{figure*}

The frequency response of the thermal generation units has been
modeled through the transfer functions proposed
in~\cite{kundur94,neplan16}. Parameters have been selected from
typical values, see Table~\ref{tab.thermal}. The combined cycle
generation unit frequency behavior is supposed similar to the gas
generation units, see Fig. \ref{fig.thermal_generation_models}. The
two inputs for these three frequency models are $(i)$ frequency
deviations ---including constraints provided by the frequency
containment---, and $(ii)$ AGC conditions for the frequency
restoration in isolated power systems. Both inputs are linked by the
corresponding droop.

\begin{table}[tbp]
  \caption{Block diagram parameters of thermal units models~\cite{kundur94,neplan16}}
  \label{tab.thermal}
  \centering
  \resizebox{0.998\linewidth}{!}{
    \begin{tabular}{cccccc} 
      \hline
      \multicolumn{2}{c}{{\bf{Gas and combined cycle units}}} & \multicolumn{2}{c}{{\bf{Diesel units}}} & \multicolumn{2}{c}{{\bf{Steam units}}}\\
      \hline
      $TR_{g}$ & 0.05 s & $T1_{d}$ & 0.01 s & $TR_{s}$ & 0.2 s\\
      $T1_{g}$ & 0.6 s & $T2_{d}$ & 0.0 s & $TSM_{s}$ & 0.1 s\\
      $T2_{g}$ & 0.5 s & $T3_{d}$ & 2.0 s & $TCH_{s}$ & 0.3 s\\
      $T3_{g}$ & 0.01 s & $T4_{d}$ & 0.1 s & $R_{s}$ & 0.05 pu\\
      $T4_{g}$ & 0.24 s & $T5_{d}$ & 0.1 s & $H_{s}$ & 5 s\\
      $TD_{g}$ & 0.2 s & $T6_{d}$ & 0.1 s &  & \\
      $R_{g}$ & 0.05 pu & $K_{d}$ & 3 &  & \\
      $R_{cc}$ & 0.05 pu & $R_{d}$ & 0.05 pu &  & \\
      $H_{g}$ & 5 s & $H_{d}$ & 2.45 s &  & \\
      $H_{cc}$ & 5 s &  &  &  & \\
      \hline
    \end{tabular}
  }
\end{table}

According to the Spanish insular power system requirements, the AGC
system is in charge of removing the steady-state frequency error after
the frequency containment control. This is usually known as `frequency
restoration', and modeled in a similar way to~\cite{perez14}. The
equivalent regulation effort $\Delta RR$ is then estimated as:
\begin{equation}\label{eq.rr}
\Delta RR = - K_{f} \cdot \Delta f.
\end{equation}

This expression is included in the block diagram shown
Fig.~\ref{fig.dynamic_model}, in block {\texttt{`AGC'}}. $K_{f}$ is
determined following the ENTSO-E recommendations~\cite{entsoe}. This
regulation effort is conducted by all thermal generation units and
distributed depending on the participation factors~$K_{u,i}$, assuming
that: $(i)$~all thermal generation units connected to the power system
equally participate in secondary regulation control and $(ii)$~the
participation factors are obtained as a function of the speed droop of
each unit. The participation factors~$K_{u,i}$ of thermal generation
units disconnected from the grid are considered as zero~\cite{wood12}:
\begin{equation}\label{eq.pref}
  \begin{split}
    \Delta P_{ref,th}=\dfrac{1}{T_{u,th}}\int\Delta RR\cdot K_{u,th}\;dt =  \\
    = \dfrac{-1}{T_{u,th}} \cdot K_{u,th} \cdot K_f \int \Delta f \;dt.
  \end{split}
\end{equation}

\begin{equation}\label{eq.Kuth}
\begin{split}
\sum K_{u,th}=\sum_{m=1}^{4}K_{u,s,m}+\sum_{q=1}^{5}K_{u,g,q} +\\
+\sum_{k=1}^{5}K_{u,d,k}+\sum_{l=1}^{2}K_{u,cc,l}=1 
\end{split}
\end{equation}

\subsubsection{Load shedding}\label{sec.load_shedding}

A realistic load shedding scheme is considered in the proposed model
by means of sudden load disconnections when frequency excursions are
higher than certain thresholds. Table~\ref{tab.load} summarizes these
frequency thresholds, time delay and load shedding values for
different scenarios. This load shedding scheme depends on the
islanding power system operation conditions required by the Spanish
TSO and thus, the responses are in line with certain frequency
excursion thresholds. When the scenario corresponds to an intermediate
load case, the load shedding value is interpolated from the
corresponding steps.

\begin{table}[tbp]
  \caption{Load shedding scheme.}
  \label{tab.load}
  \centering
  \resizebox{.998\linewidth}{!}{%
    \begin{tabular}{ccccc}
      \hline
      \multirow{ 2}{*}{{\bf{Step}}} & \multirow{ 2}{*}{{\bf{Threshold (Hz)}}} & \multirow{2}{*}{{\bf{Delay (s)}}} & \multicolumn{2}{c}{{\bf{Load shedding (MW)}}}\\
                                    & & & {\em{Peak demand}} & {\em{Valley demand}} \\
      \hline
      1 & 48.9 & 0.1 & 14.6 & 5.8\\
      2 & 48.9 & 0.2 & 16.2 & 7.0\\
      3 & 48.8 & 0.4 & 17.1 & 8.6\\
      4 & 48.8 & 0.6 & 41.1 & 18.8\\
      5 & 48.5 & 0.1 & 8.0 & 4.1\\
      6 & 48.5 & 0.2 & 27.3 & 11.8\\
      7 & 48.4 & 0.4 & 17.5 & 7.7\\
      8 & 48.1 & 0.1 & 17.9 & 9.7\\
      \hline
    \end{tabular}
  }
\end{table}

\subsubsection{Wind power plants}\label{sec.wind_turbines} 

One equivalent VSWT with $n_{WT}$-times the size of each one model the
wind power penetration ---being $n_{WT}$ the number of wind
turbines~\cite{pyller03,mokhtari14}---, is proposed as aggregated
model for wind power plants. An equivalent averaged wind speed
($v_{w}=10.25$~m/s) is assumed for the simulations. This wind speed is
considered as constant, which has been previously used in the specific
literature for short-time period frequency analysis including wind
power
plants~\cite{chang10,erlich10,margaris12,zertek12,vzertek12}. With
this wind speed, the wind generation accounts for 80\% of the
installed wind power capacity.

\begin{figure*}[tbp]
  \centering
  \subfloat[Frequency control strategy]{\includegraphics[width=.4\linewidth]{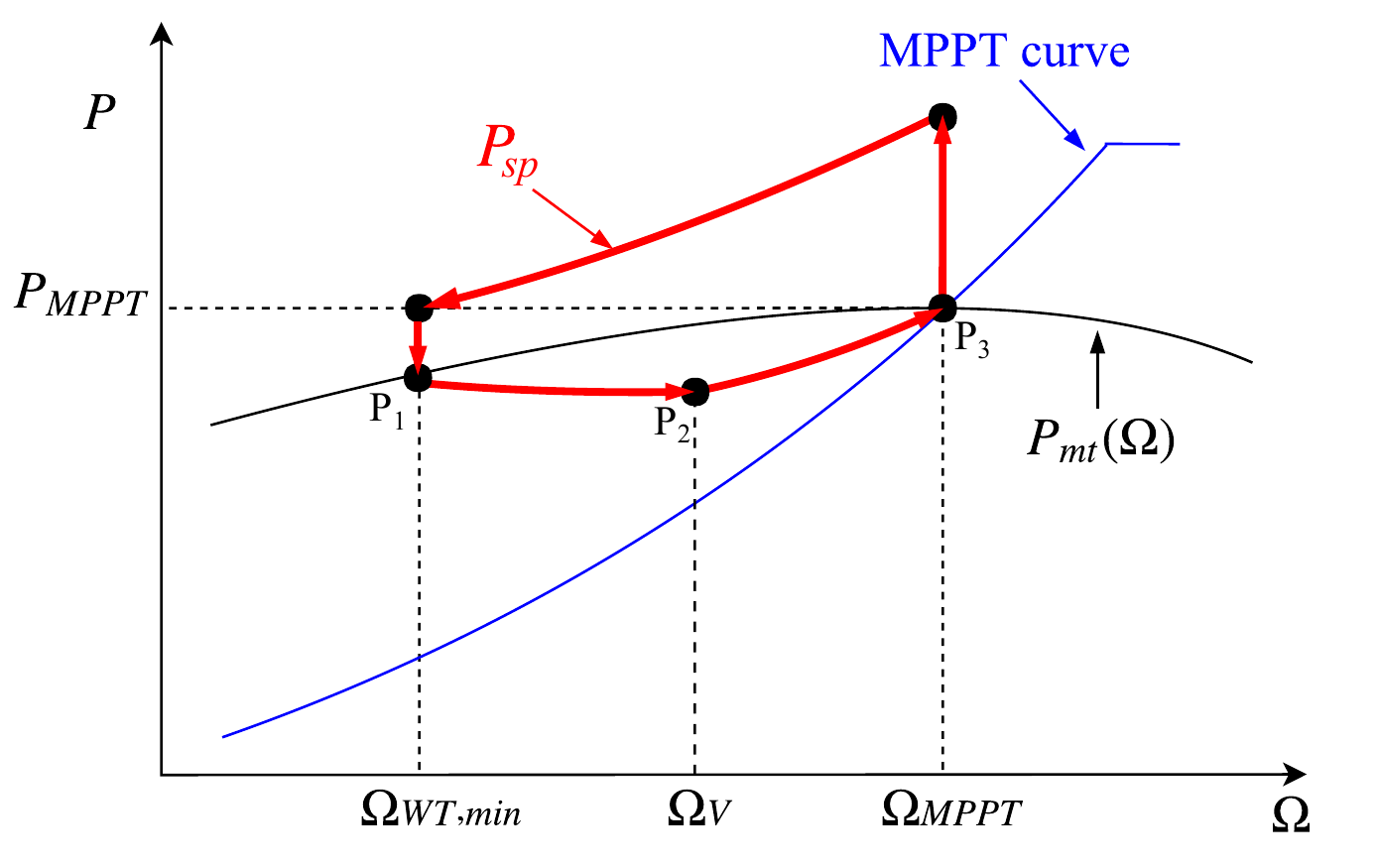}\label{fig.control}}
  \subfloat[$\Delta P_{WF}$ with frequency control strategy]{\includegraphics[width=.4\linewidth]{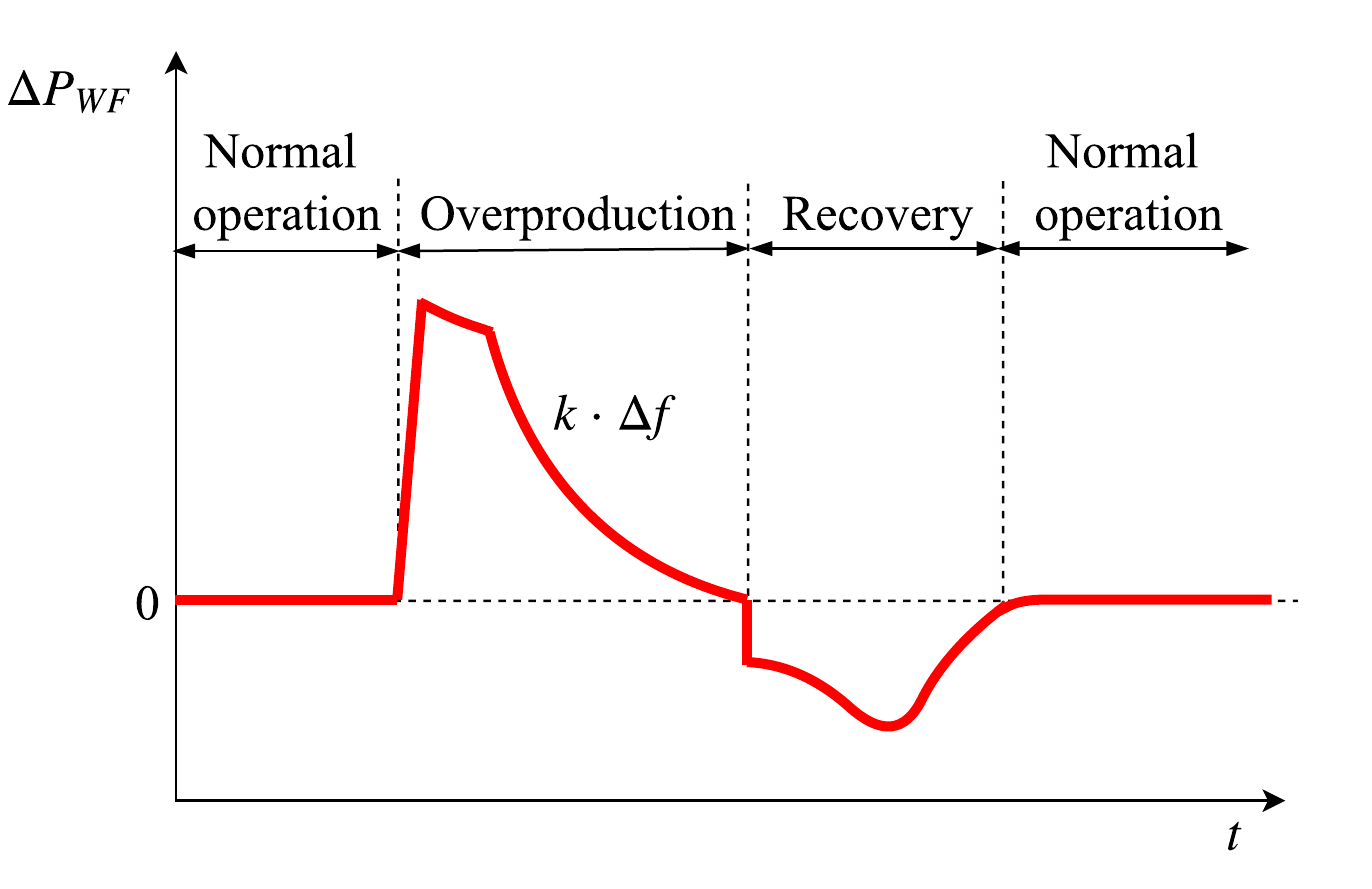}\label{fig.power}}
  \caption{Wind frequency control strategy and VSWTs active power variation \cite{fernandez18}.}
  \label{fig.tfcf}
\end{figure*}

Wind turbines are modeled according to the turbine control model,
mechanical two-mass model and wind power model described
in~\cite{ullah08,clark10}. The two-mass model assumes the rotor and
blades as a single mass, and the generator as another
mass~\cite{jafari17,liu17}. Huerta \emph{et al.} consider that the
two-mass model is the most suitable to evaluate the grid
stability~\cite{huerta17}. Wind turbines also include a frequency
control response. The strategy for VSWTs implemented in this paper is
based on the technique described
in~\cite{fernandez18,fernandez18fast}, see
Fig.~\ref{fig.control}. It was evaluated in~\cite{fernandez18} for
isolated power systems with up to a 45\% of wind power integration and
compared to the approach of~\cite{tarnowski09}, providing a more
appropriate frequency response under power imbalance
conditions. In~\cite{fernandez18fast}, the proposed frequency control
strategy was studied for multi-area power systems. Three operation
modes are considered: $(i)$~normal operation mode,
$(ii)$~overproduction mode and $(iii)$~recovery mode. Different
set-point active power $P_{sp}$ values are then determined aiming to
restore the grid frequency under power imbalance
conditions. Fig.~\ref{fig.power} depicts the VSWTs active power
variations $\Delta P _{WF}$ submitted to an under-frequency excursion,
being $\Delta P_{w}=P_{sp}-P_{MPPT}(\Omega_{MPPT})$.

\begin{figure*}[tbp]
  \centering
  \includegraphics[width=0.75\linewidth]{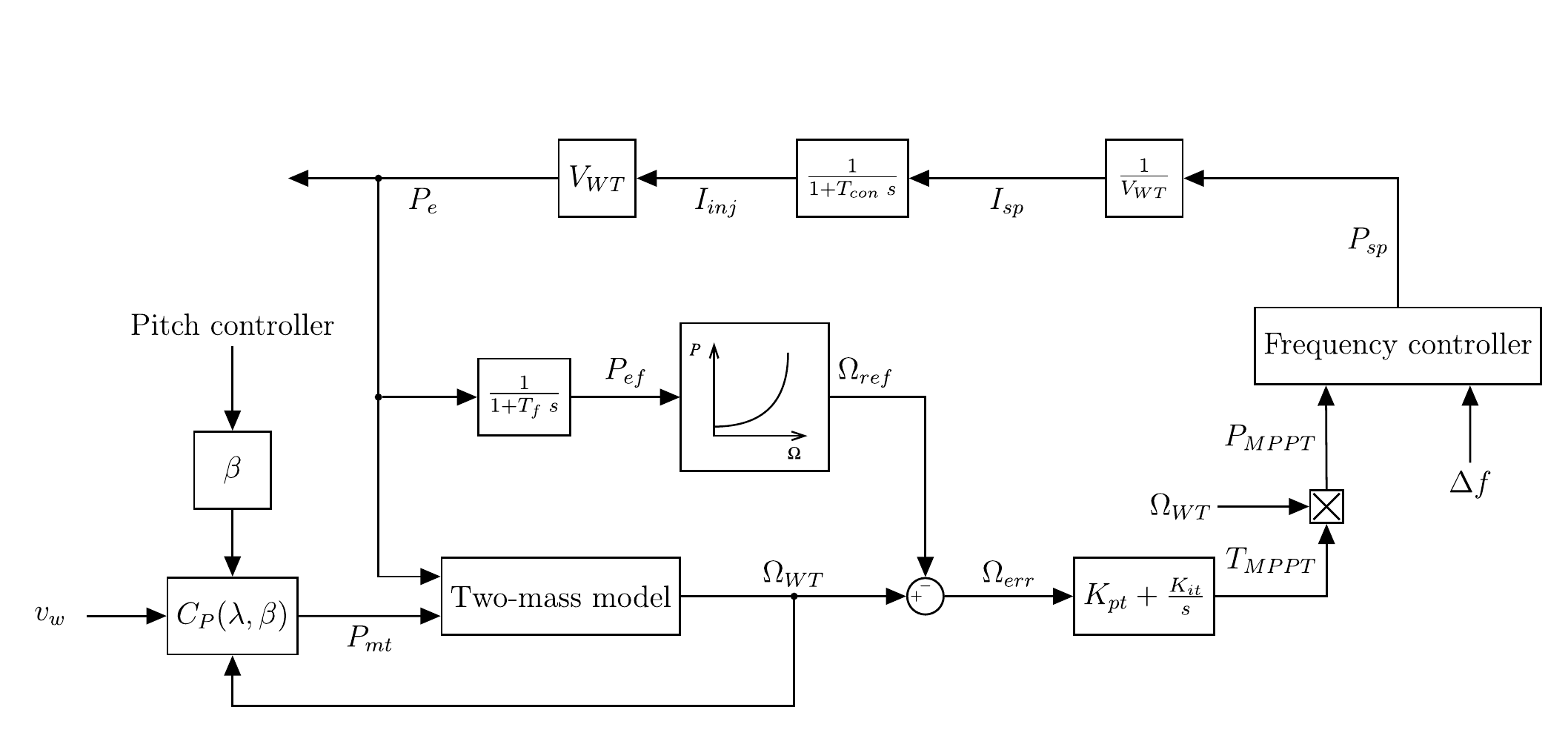}
  \caption{Variable speed wind turbine model with frequency controller}
  \label{fig.aero_control}
\end{figure*}

\begin{figure*}[tbp]
  \centering
  \subfloat[Frequency evolution]{\includegraphics[width=0.35\linewidth]{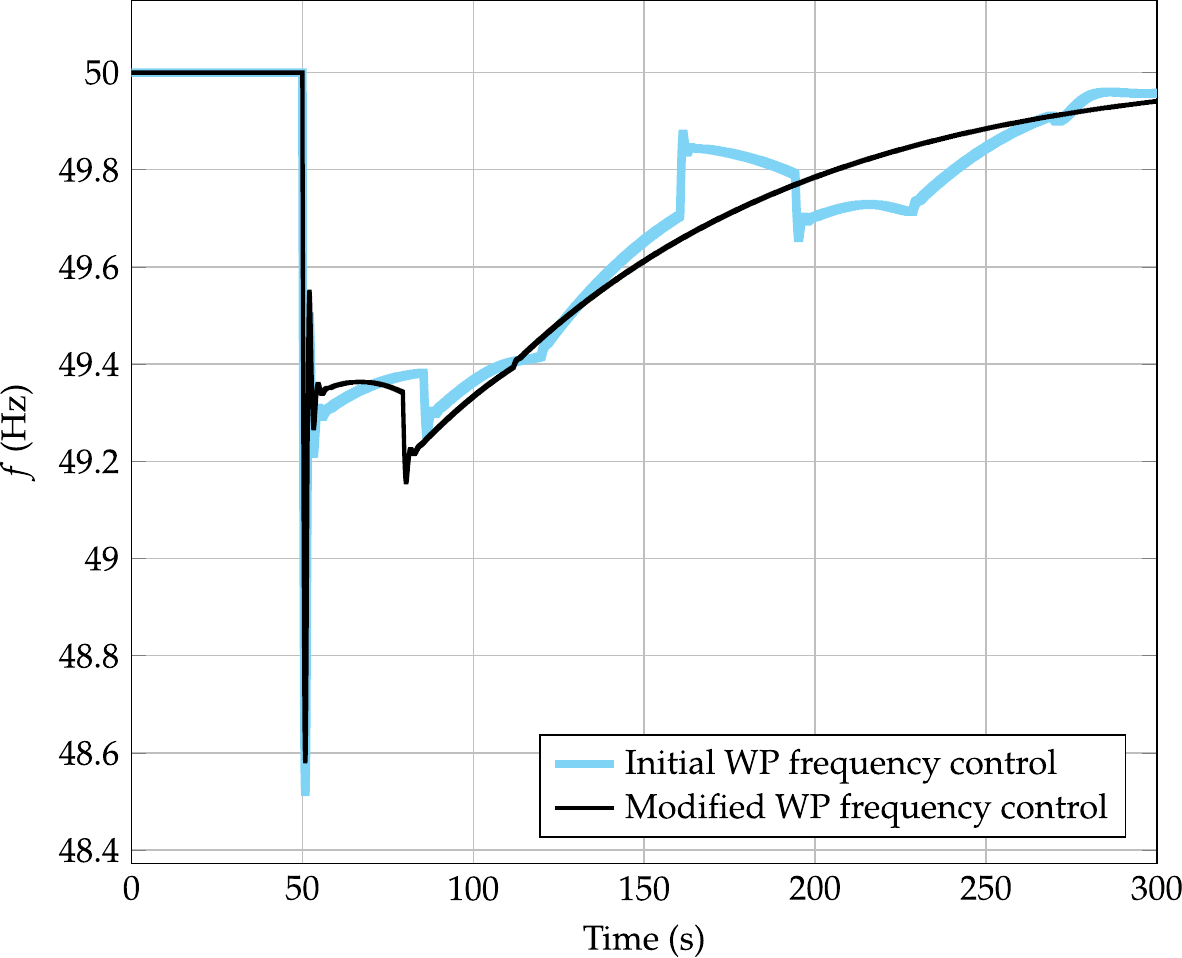}}
  \subfloat[Wind power generation]{\includegraphics[width=0.35\linewidth]{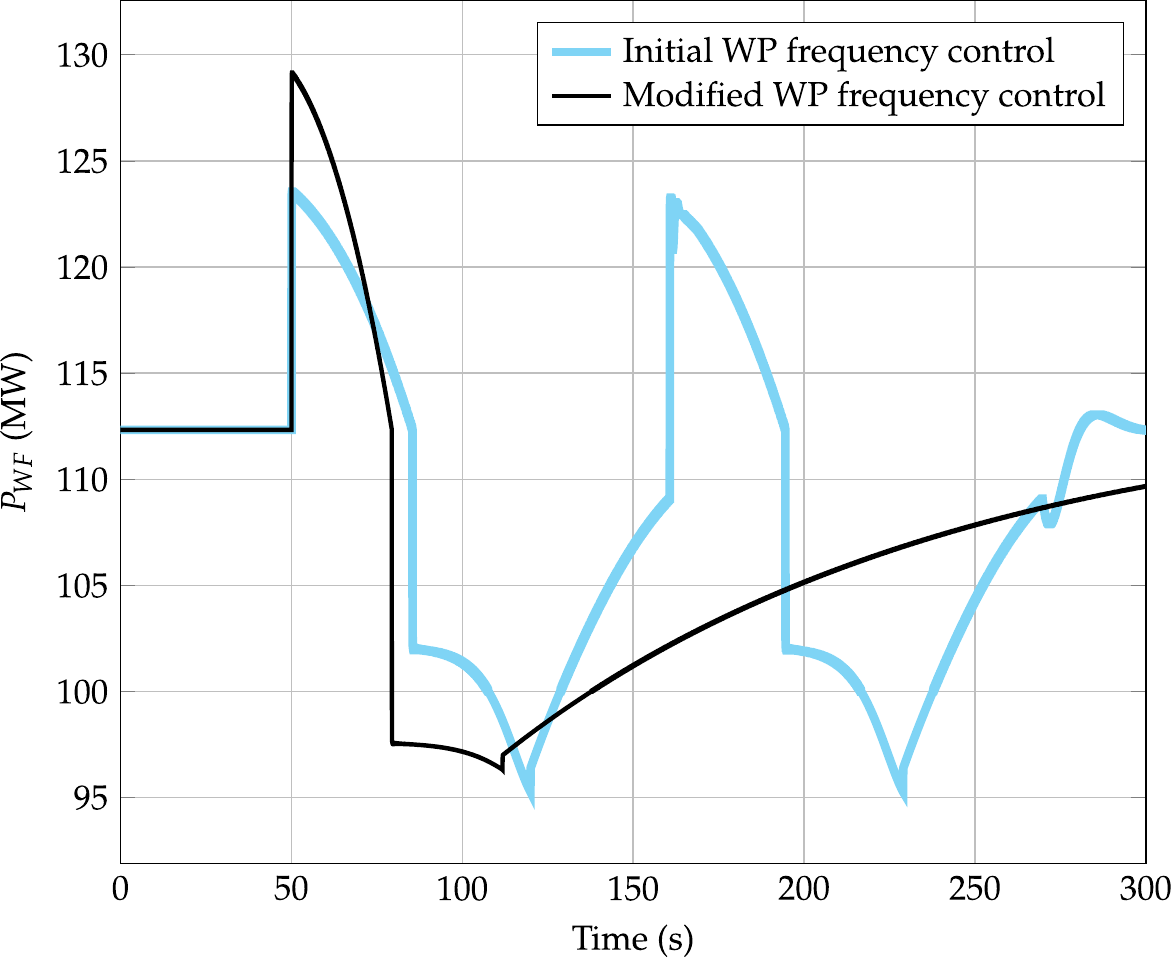}}
  \caption{Frequency deviation and wind power generation when using the original wind frequency controller}
  \label{fig.potencia}
\end{figure*}

With the aim of reducing load shedding contributions under high wind
power integration scenarios, two modifications are carried out to the
preliminary frequency controller, both in overproduction and recovery
periods. According to~\cite{fernandez18}, the overproduction
power~$\Delta P_{OP}$ is estimated proportionally to the frequency
excursion evolution, with a maximum value of 10\%. In this paper, the
maximum $\Delta P_{OP}$ is increased to 15\%, in order to provide more
power after the imbalance and minimizing load shedding situations. In
the recovery mode, the power of point $P_{2}$ is defined as
$P_{MPPT} (\Omega_{V})+x\cdot \left( P_{mt}(\Omega_{V})-P_{MPPT}
  (\Omega_{V})\right) $ ---see Fig.~\ref{fig.control}---, being $x$
an scale factor considered as 0.75 in the original
approach~\cite{fernandez18}. However, in this case, the recovery time
period of the wind turbines is faster than the AGC action of the
frequency restoration control, and subsequently obtaining an
undesirable frequency evolution, see Fig.~\ref{fig.potencia}. As a
consequence, $x$ has been increased to 0.95, smoothing and slowing
down the recovery period of the wind frequency controller. This
alternative frequency controller is included in the VSWT model as seen
in Fig.~\ref{fig.aero_control}.

\section{Results}\label{sec.results}

\subsection{Scenarios under consideration}

\begin{figure}[tbp]
	\centering
	\includegraphics[width=0.995\linewidth]{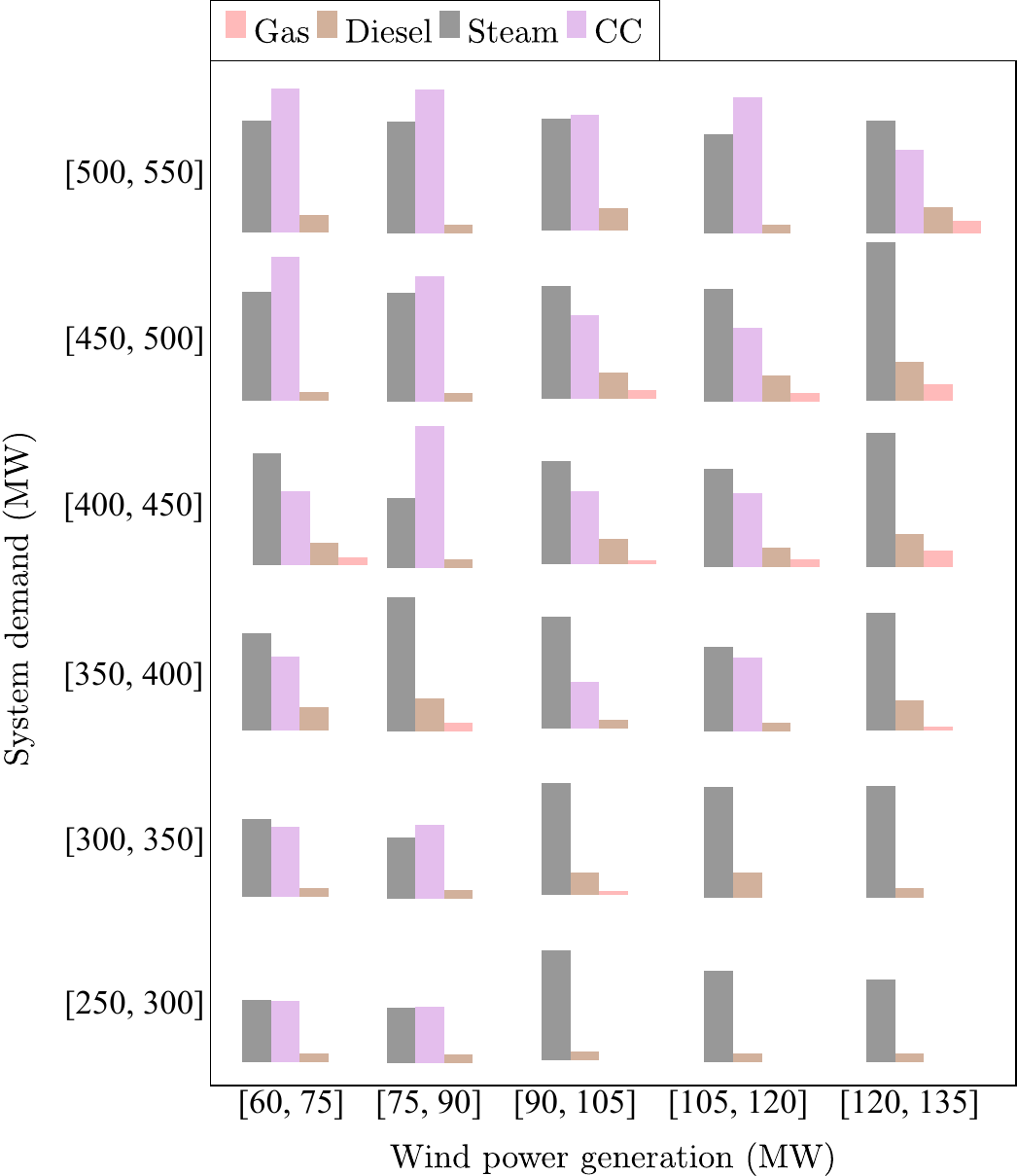}
	\caption{Scenarios under study}
	\label{fig.scenarios2}
\end{figure}

\begin{figure}[tbp]
	\centering
	\includegraphics[width=0.995\linewidth]{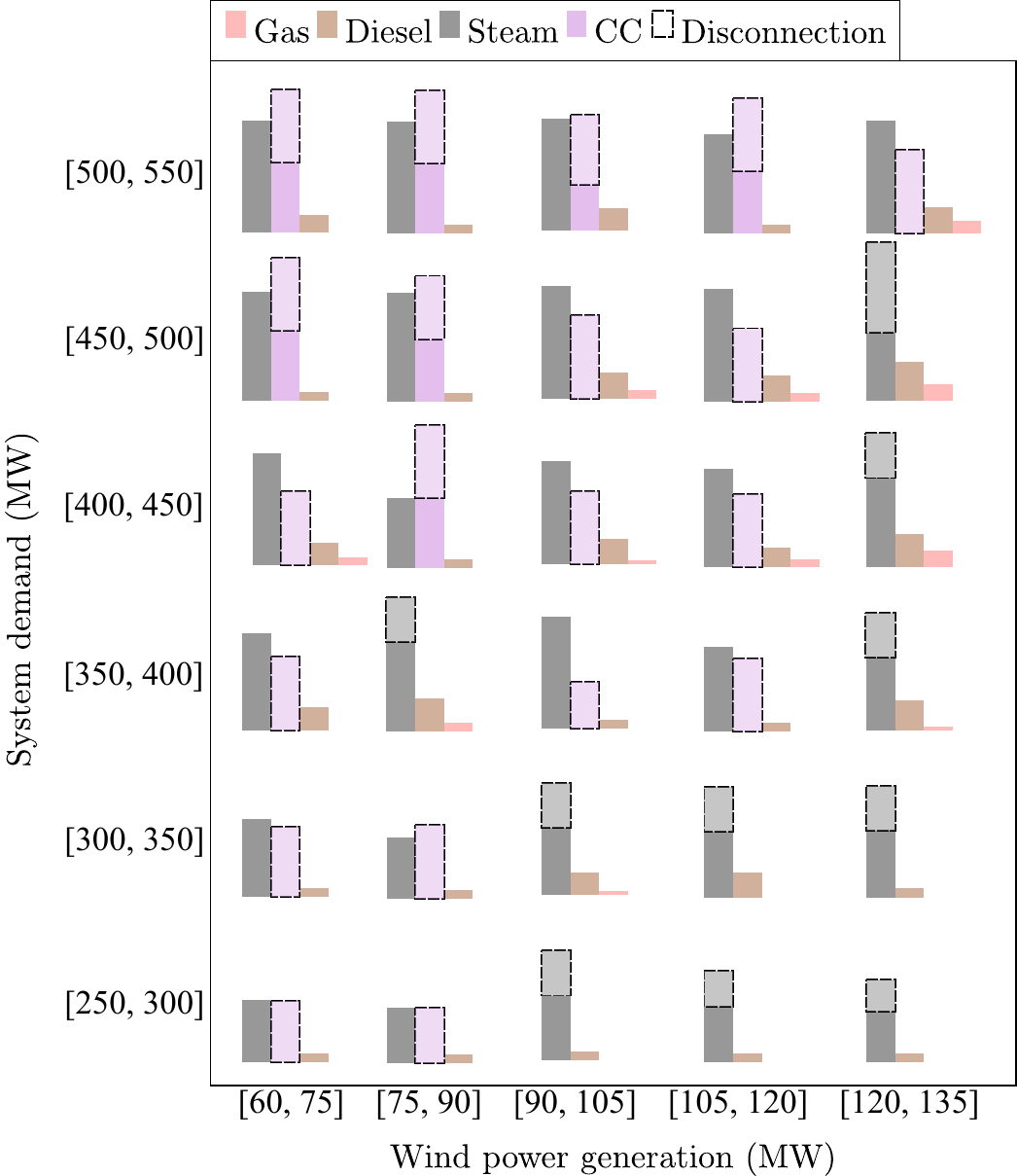}
	\caption{Generation mix after disconnections}
	\label{fig.after_imbalance}
\end{figure}

According to the demand distribution in Gran Canaria Island along 2018
previously discussed in Section
{\ref{sec.power_system_general_overview}}, six different power demand
conditions are considered for the study. Each system demand is
analyzed under different wind power generation percentages following
Fig.~\ref{fig.demand}. Thus, thirty different energy scenarios are
under study, which is significantly higher than other contributions
focused on frequency control under contingencies including wind power
plants \cite{lalor2005frequency,sigrist2009representative}.  To
determine the energy schedule of each supply-demand scenario, the Unit
Commitment model described in Section \ref{sec.optimization_model} is
run with GAMS software and Cplex 12.2 solver, which uses a branch and
cut algorithm to solve MILP problems. Fig.~\ref{fig.scenarios2}
depicts the energy schedule of each scenario aggregated by generation
technology.  From the $N-1$ criterion, the largest generation unit is
suddenly disconnected under a contingency. As a consequence, a
different generation group is disconnected in each scenario, depending
on the energy schedule obtained by the UC model and subsequently
addressed a variety of power imbalance
situations. Fig.~\ref{fig.after_imbalance} summarizes the energy
schedule of each scenario after these disconnections, pointed out the
technology and generation unit tripping under such circumstances. Due
to these sudden disconnections, the equivalent rotational inertia of
the power system is reduced according to eq. ({\ref{eq.Tm}}).

\subsection{Frequency response analysis}\label{sec.simulation_results}

\begin{figure*}[tbp]
  \centering
  \subfloat[Without wind frequency response]{\includegraphics[width=.355\linewidth]{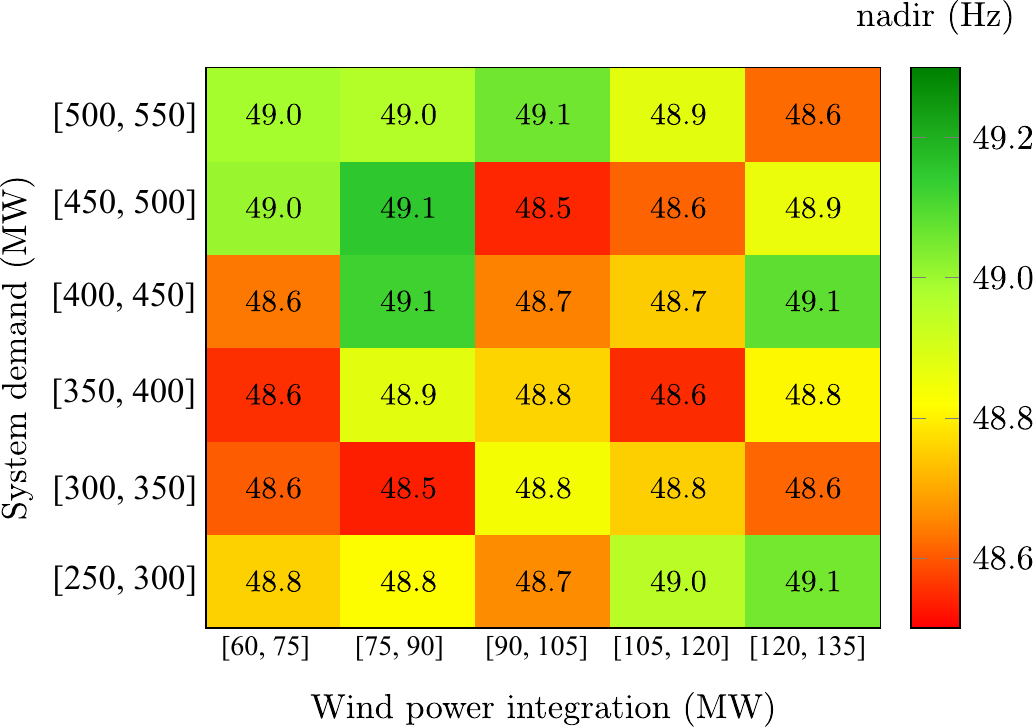}\label{fig.nadir_sin_2}}
  \qquad\qquad\qquad
  \subfloat[Without wind frequency response and $\Delta P=10$\%]{\includegraphics[width=.355\linewidth]{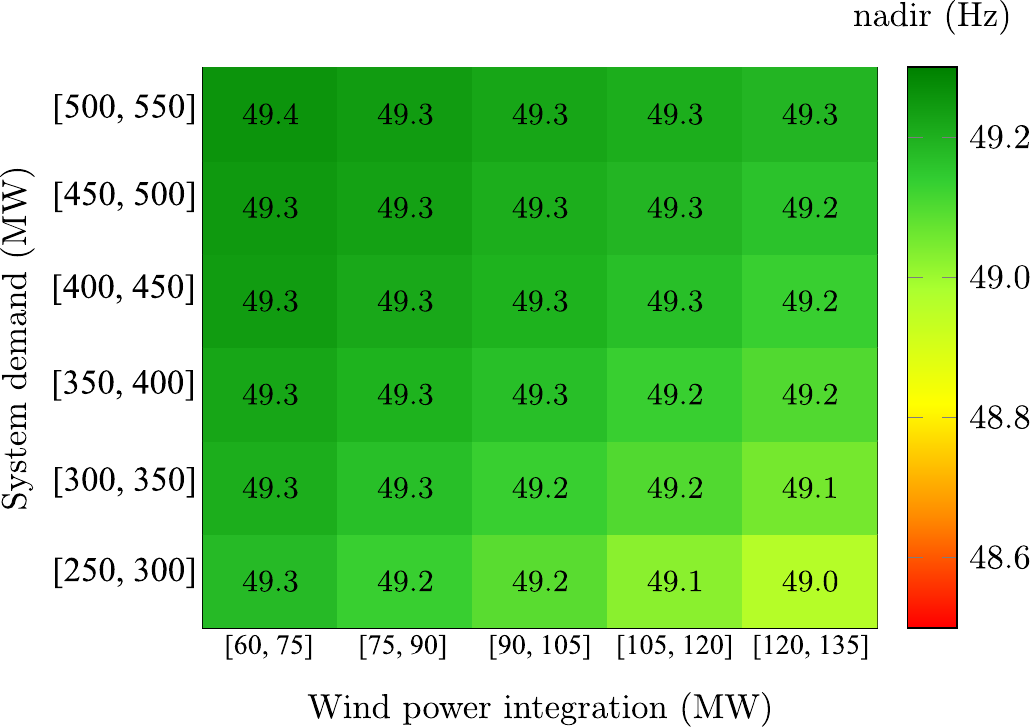}\label{fig.nadir_sin}}
  \\
  \subfloat[With wind frequency response]{\includegraphics[width=.355\linewidth]{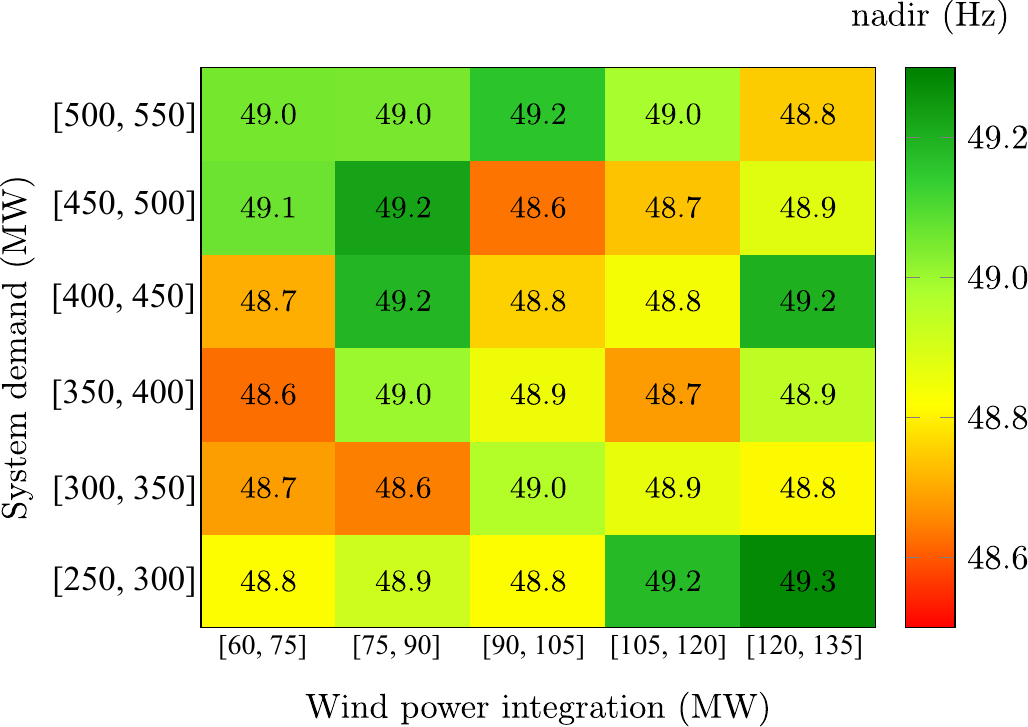}\label{fig.nadir_con_2}}
  \qquad\qquad\qquad
  \subfloat[With wind frequency response and $\Delta P=10$\%]{\includegraphics[width=.355\linewidth]{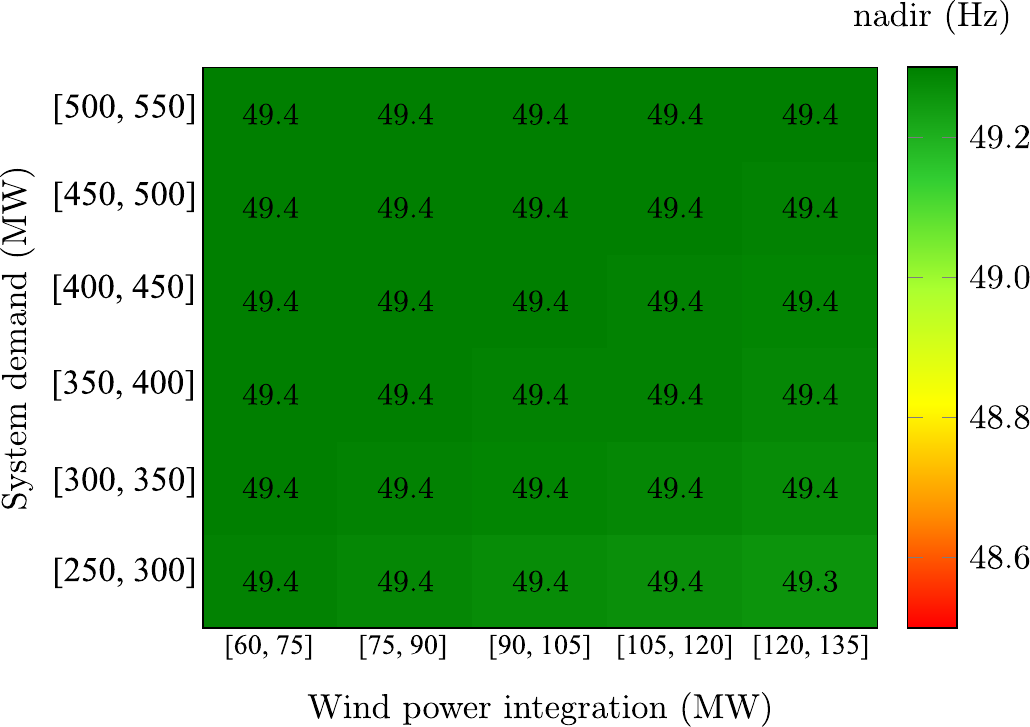}\label{fig.nadir_con}}
  \caption{\textcolor{black}{Nadir estimation: power demand and wind power integration in Gran Canaria}}
  \label{fig.nadir}
\end{figure*}

\begin{figure*}[tbp]
  \centering
  \subfloat[Without wind frequency response]{\includegraphics[width=.35\linewidth]{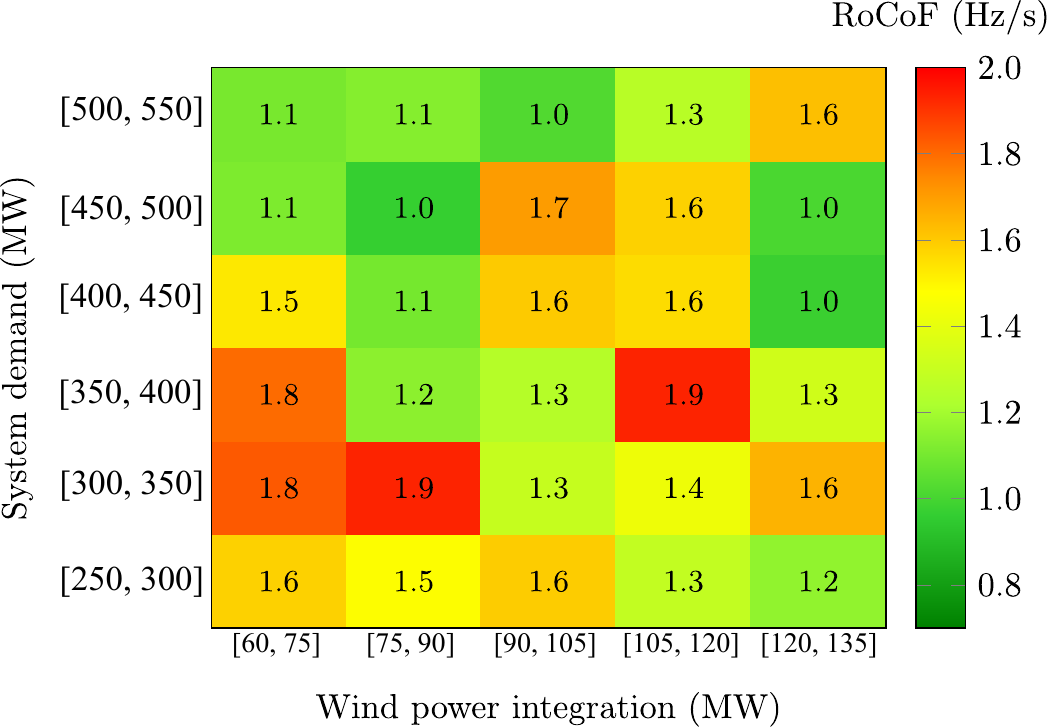}\label{fig.rocof_sin_2}}
  \qquad\qquad\qquad
  \subfloat[Without wind frequency response and $\Delta P=10$\%]{\includegraphics[width=.35\linewidth]{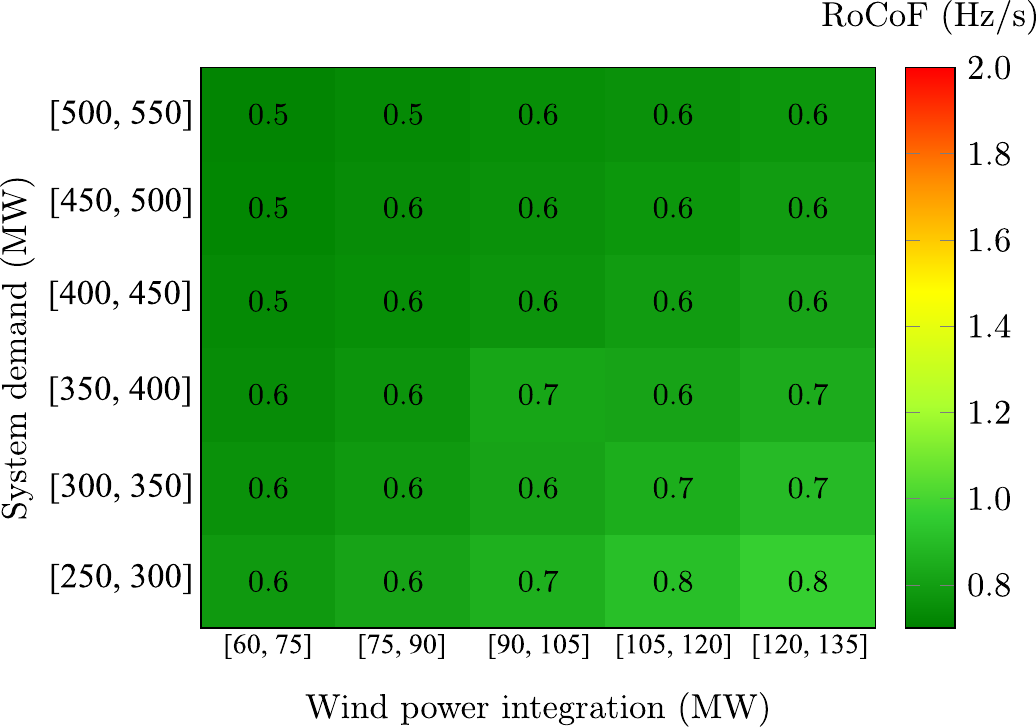}\label{fig.rocof_sin}}\\
  \subfloat[With wind frequency response]{\includegraphics[width=.35\linewidth]{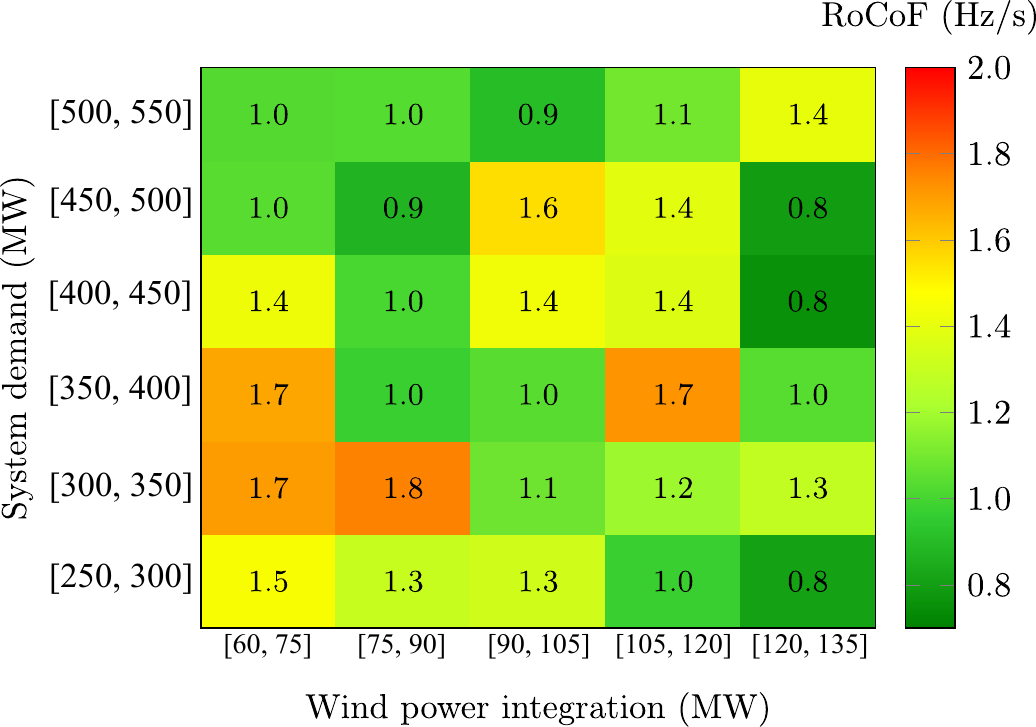}\label{fig.rocof_con_2}}
  \qquad\qquad\qquad
  \subfloat[With wind frequency response and $\Delta P=10$\%]{\includegraphics[width=.35\linewidth]{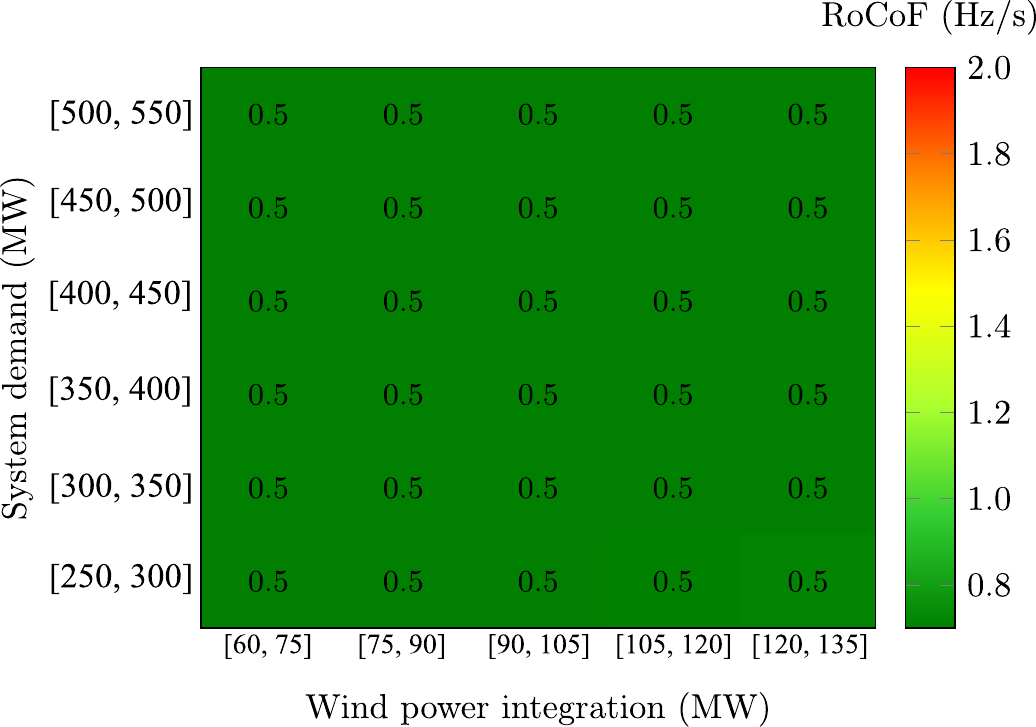}\label{fig.rocof_con}}
  \caption{\textcolor{black}{RoCoF estimation: power demand and wind power integration in Gran Canaria}}
  \label{fig.rocof}
\end{figure*}

With the aim of evaluating frequency deviation and power system
performance under the sudden generation disconnection established with
the $N-1$ criterion, grid frequency response is analyzed $(i)$
excluding wind frequency control and only considering conventional
units; and $(ii)$ including conventional units and wind frequency
control strategy. Firstly, nadir and RoCoF results for the 30
simulated scenarios according to the generation unit tripping obtained
for the UC model and depicted in Fig.~\ref{fig.after_imbalance} are
compared to results obtained following methodologies of previous
contributions~\cite{keung09,ma10,el11,alsharafi18}. They usually
assume a constant 10\% imbalance, neglect any inertia power system
modification and do not include load shedding scheme in their
frequency analysis models. With this aim, Fig.s~\ref{fig.nadir} and
\ref{fig.rocof} summarize nadir and RoCoF respectively, including (or
not) wind frequency response. RoCoF is calculated between 0.3 and
0.5~s after the sudden disconnection of the largest conventional
generation unit for each energy scenario. As can be seen, clear
differences are identified between both approaches. In fact, most
obvious results are determined with a constant power imbalance, as was
to be expected, see Fig. \ref{fig.nadir_sin} and \ref{fig.nadir_con}
for nadir comparison values. If a simplified power system modeling is
considered for frequency control analysis, with typical 10\% power
imbalance conditions ---usually assumed in previous contributions as
was previously discussed--- 49.4 Hz nadir and 0.5 Hz/s RoCoF values
are obtained for all cases, which provides significant discrepancies
with our proposal, see Fig. \ref{fig.nadir} and \ref{fig.rocof}
respectively. Indeed, nadir lies in between 48.54 and 49.15~Hz when
wind power plants are excluded from frequency control, depending on
each scenario ---see Fig. {\ref{fig.nadir_sin_2}}---. In fact, these
values were even worse if the load shedding program was not
considered, as it is activated in 21 of the 30 scenarios
analyzed. However, a larger wind power integration without frequency
control ---see Fig. {\ref{fig.nadir_sin_2}}--- doesn't imply a worse
nadir response, which could be deduced a priori, due to the loss of
the larger power plant (which is different, depending on the
scenario). When wind frequency control is considered for simulations,
the minimum frequency is increased 110~mHz in average for all
cases. Moreover, the more wind power integration providing frequency
control, the lower nadir is obtained. For instance, for wind power
integration over 50\%, the minimum frequency is reduced around
200~mHz. It can't be then deduced an homogeneous response of the
considered power system submitted to realistic generation unit
tripping. In this way, and based on the proposed methodology and
modeling, it is important to point out that higher wind power
integration excluding frequency control does not always imply a worse
frequency response, see Fig. \ref{fig.nadir_sin_2}.  With regard to
RoCoF, it varies between 0.97 and 1.93~Hz/s initially, see Fig.
{\ref{fig.rocof_sin_2}}, but slighted 185~mHz/s in average by
including wind frequency control, Fig.
{\ref{fig.rocof_con_2}}. These results are substantially different
from those obtained in the simplified power system analysis (where
RoCoF was around 0.5 Hz/s); this is mainly due to the inertia change
considered in this study and neglected in the previous one, as low
inertia is related to a faster ROCOF~\cite{daly15}. Therefore,
including wind power frequency control can lead to lower frequency
deviations under imbalances, as was
expected.  

\begin{figure}[tbp]
  \centering
  \subfloat[Load shedding without wind power frequency response]{\includegraphics[width=.98\linewidth]{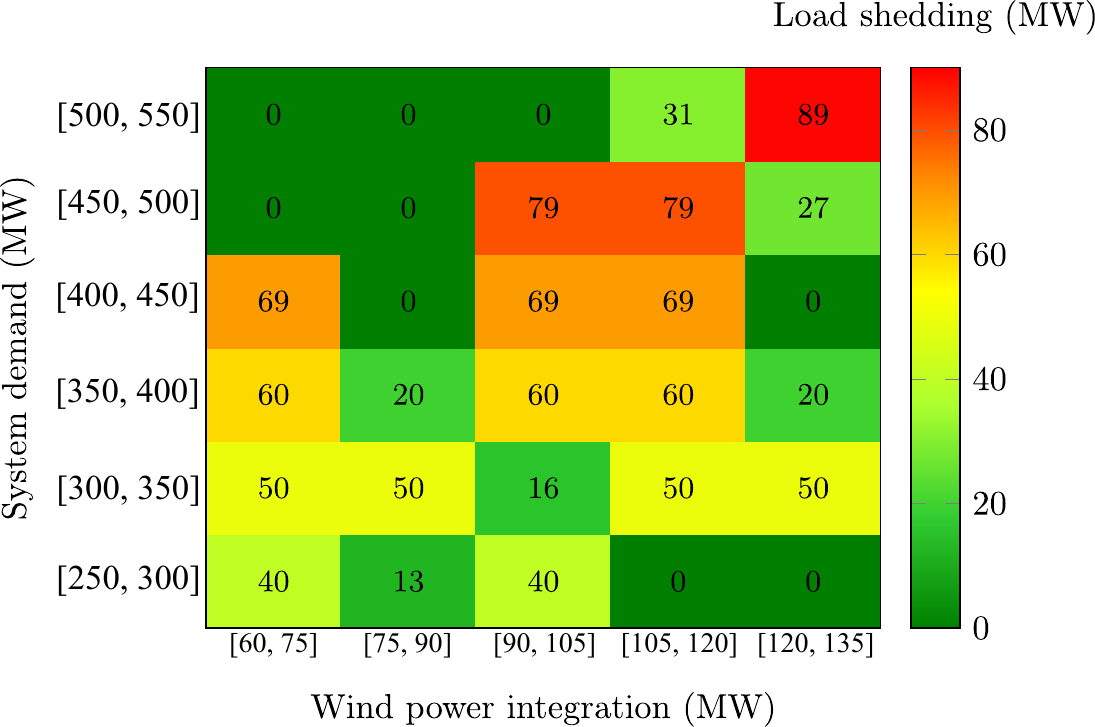}\label{fig.deslastre_sin_2}}\\
  \subfloat[Load shedding with wind power frequency response]{\includegraphics[width=.98\linewidth]{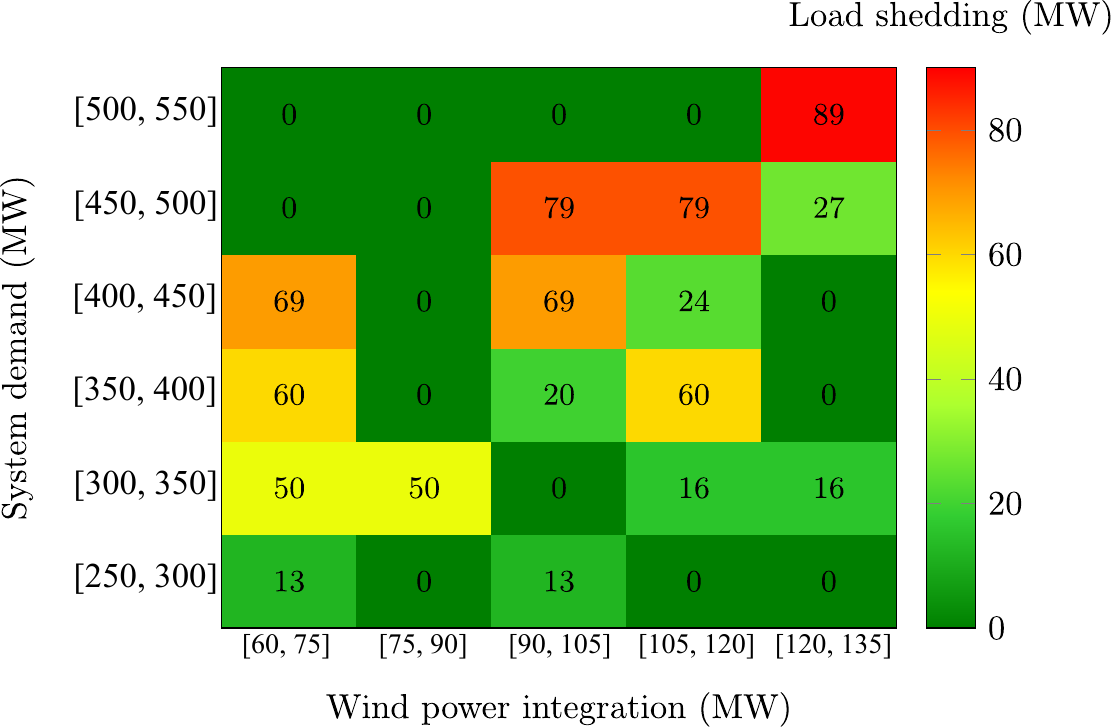}\label{fig.deslastre_con_2}}
  \caption{Load shedding: power demand and wind power integration in Gran Canaria}
  \label{fig.deslastre_res}
\end{figure}

The proposed wind frequency response analysis allows us to evaluate
the wind frequency control impact on load shedding actions in
islanding power systems under different
imbalances. Fig.~\ref{fig.deslastre_res} summarizes the load
shedding for the 30 simulated scenarios and considering the generation
unit tripping obtained for the UC model, see
Fig.~\ref{fig.after_imbalance}. In this way,
Fig.~{\ref{fig.deslastre_sin_2}} and Fig.
{\ref{fig.deslastre_con_2}} shows the corresponding load shedding
responses by including or not wind frequency control for the
considered energy scenarios. Both nadir and RoCoF improvements lead to
a load shedding reduction in 11 scenarios. Moreover, in these 11
scenarios, the average load shedding reduction is 80\%, getting up to
a 100\% reduction in 5 scenarios ---for example, compare 30.80 MW load
shedding under the range [500, 550]~MW power demand and [105, 120]~MW
wind power generation, Fig.~{\ref{fig.deslastre_sin_2}}, to 0 MW
load shedding under the same demand and wind power values when wind
frequency control is included, Fig. {\ref{fig.deslastre_con_2}}---.

Table \ref{tab.overview} shows a comparison of results between the
proposed analysis described in this paper and conventional
methodologies previously considered where a constant imbalance is
assumed, inertia of the power system is kept constant during the
imbalance and load shedding is not included for simulations. In the
table, the average $\mu$ and variance $\sigma^{2}$ of nadir, RoCoF
inertia change and load shedding values for the 30 different
generation mix and imbalance scenarios are shown with and without wind
frequency control.

\begin{table*}[tbp]
  \caption{\textcolor{black}{Comparison of results: nadir, RoCoF, inertia change and load shedding}}
  \label{tab.overview}
  \centering
  \resizebox{0.85\linewidth}{!}{
    \begin{tabular}{cccccc} 
      \hline
      &&\multicolumn{2}{c}{Without wind frequency control} & \multicolumn{2}{c}{With wind frequency control} \\
      &&\textbf{$\mu$} & \textbf{$\sigma^{2}$}&\textbf{$\mu$} & \textbf{$\sigma^{2}$}\\
      \hline
      \multirow{4}{*}{Proposed analysis} & nadir (Hz) & 48.8 & 0.036 & 48.9 & 0.037 \\
      & RoCoF (Hz/s) & 1.4 & 0.085 & 1.2 & 0.085 \\
      & Inertia change (s) & 1.9 & 0.64 & 1.9 & 0.64 \\
      & Load shedding (MW) & 34.9 & 894.7 & 24.7 & 938.6 \\\hline
      \multirow{4}{*}{Previous approaches} & nadir (Hz) & 49.2 & 0.006 & 49.4 & 0.0002 \\
      & RoCoF (Hz/s) & 0.6 & 0.005 & 0.5 & 0.0002 \\
      & Inertia change (s) & --- & --- & --- & ---\\
      & Load shedding (MW) & --- & --- & --- & --- \\
      \hline
    \end{tabular}
  }
\end{table*}

\section{Conclusion} \label{sec.conclusions}

{\textcolor{black}{Frequency excursions are analyzed in an isolated
    power system by considering the loss of the largest conventional
    generation group, including wind frequency control strategy, load
    shedding and energy scenarios obtained by a UC model.}} The case
study is focused on the real isolated power system located in the Gran
Canaria Island (Spain), which has doubled its wind power capacity in
the last two years. With regard to the frequency analysis, by
including wind power generation into frequency control, nadir and
RoCoF are reduced in most of energy scenarios considered (110~mHz and
185~mHz/s in average, respectively). Regarding load shedding, it is
reduced in 11 out of the 30 power imbalance analyzed. This improvement
is more significant in high wind power integration scenarios
(regardless of the power demand), and for high power demands
(regardless of the wind power integration). Therefore, wind frequency
control can be considered a remarkable solution to reduce load
shedding in islanding power systems with high wind power integration.

{\textcolor{black}{It can be affirmed that there is no homogeneity in
    the frequency response results, providing a clear dependence on
    the dispatch of the conventional rotational generation units,
    their participation in the global energy scenario and the inertia
    changes addressed by the generation unit tripping ---for example
    $\mu (\Delta f) = 1.2$ Hz and $\sigma^{2} (\Delta f ) = 0.036$ for
    the nadir results. Subsequently, the participation of wind power
    plants into frequency control should be analyzed by considering
    not only the wind power integration level, but also other
    variables, such as the energy scenario, the rate power of the
    different generation units, as well as the reserves provided by
    conventional generation units and thus, the equivalent rotational
    inertia given by them; aspects that have not explicitly taken into
    account in previous works. Indeed, nadir and RoCoF values give
    almost homogeneous values when such aspects are not included in
    the simulations, with a typical deviation near zero.}}
{\textcolor{black}{In addition, and according to the remarkable
    frequency excursion dependence on different parameters such as on
    power system rotational inertia, generation unit technologies and
    rate power of the tripping-generation unit; controller parameters
    should dynamically vary in line with those power system variables
    and at the same time keeping frequency requirements. These aspects
    are currently under analysis by the authors for further
    contributions. Whereas our findings are derived based on a
    single-islanding power system case study, our results are useful
    and significant based on the relevance of our realistic case study
    coupled with the number of different energy scenarios and wind
    power integration ranges.}}

\section{Acknowledgment}

Authors thank Ignacio Ares for the preliminary analyses that he did
as part of his final master project.

\section{Funding}

This work has been partially supported the Spanish Ministry of Economy
and Competitiveness under the project `Value of pumped-hydro energy
storage in isolated power systems with high wind power penetration' of
the National Plan for Scientific and Technical Research and Innovation
2013-2016 (Ref.~ENE2016-77951-R) and by the Spanish Education, Culture
and Sports Ministry (Ref.~FPU16/04282).

\bibliography{biblio}

\end{document}